\documentclass[twocolumn,bbm, twocolappendix, trackchanges]{aastex631} 

\shorttitle{Halo spins of barred and unbarred galaxies}
\shortauthors{Ansar \& Das}
\graphicspath{{./}{figures/}}

\begin{document}

\title{The stellar bar -- dark matter halo connection in the TNG50 simulations}

\correspondingauthor{Sioree Ansar}
\email{sioreeansar@gmail.com}

\author[0000-0002-3353-6421]{Sioree Ansar}
\affiliation{Indian Institute of Astrophysics, Bangalore 560034, India}
\affiliation{Pondicherry University, R.V. Nagar, Kalapet 605014, Puducherry, India }

\author[0000-0001-8996-6474]{Mousumi Das}
\affiliation{Indian Institute of Astrophysics, Bangalore 560034, India}

\begin{abstract}
Stellar bars in disk galaxies grow as stars in near circular orbits lose angular momentum to their environments, including their Dark Matter (DM) halo, and transform into elongated bar orbits. This angular momentum exchange during galaxy evolution hints at a connection between bar properties and the DM halo spin $\lambda$, the dimensionless form of DM angular momentum. We investigate the connection between halo spin $\lambda$ and galaxy properties in the presence/absence of stellar bars, using the cosmological magneto-hydrodynamic TNG50 simulations at multiple redshifts ($0<z_r<1$). We determine the bar strength (or bar amplitude, $A_2/A_0$), using Fourier decomposition of the face-on stellar density distribution. We determine the halo spin for barred and unbarred galaxies ($0<A_2/A_0<0.7$) in the centre of the DM halo, close to the galaxy's stellar disk. At $z_r=0$, there is an anti-correlation between halo spin and bar strength. Strongly barred galaxies ($A_2/A_0>0.4$) reside in DM halos with low spin and low specific angular momentum at their centers. In contrast, unbarred/weakly barred galaxies ($A_2/A_0<0.2$) exist in halos with higher central spin and higher specific angular momentum. The anti-correlation is due to the barred galaxies' higher DM mass and lower angular momentum than the unbarred galaxies at $z_r=0$, as a result of galaxy evolution. At high redshifts ($z_r=1$), all galaxies have higher halo spin compared to those at lower redshifts ($z_r=0$), with a weak anti-correlation for galaxies having $A_2/A_0> 0.2$. The formation of DM bars in strongly barred systems highlights how angular momentum transfer to the halo can influence its central spin. 

\end{abstract}

\keywords{galaxies: haloes -- galaxies: bar -- cosmology: dark matter -- software: simulations}

\section{Introduction} \label{sec:intro}

Barred galaxies constitute a major fraction of the disk galaxy population in our local Universe. Studies using controlled galaxy simulations and cosmological simulations have shown that DM halo plays an important role in the formation of bar instability and in the evolution of bar in the stellar disk of galaxies \citep{Efstathiou.et.al.1982, Ostriker.and.Peebles.1973, Marioni.et.al.2022, Ansar.et.al.2023b}. 

The study of bar and DM halo interaction using isolated controlled galaxy simulations has the advantage of tracking specific dynamical processes, for example, angular momentum transfer between the disk and the halo (see \citet{Sellwood.2014} for a detailed review). Stars in circular orbits in stellar disks shift into bar orbits by losing angular momentum to the environment (including the surrounding DM halo) through resonance interactions \citep{Lynden-Bell1972, Athanassoula.2002, Petersen2016, Collier.et.al.2019a, Collier.et.al.2019b,Petersen2019, Li.et.al.2023a, Li.et.al.2023b}. As additional stellar orbits lose angular momentum and begin moving along the bar, the bar gains strength, lengthens, and its angular velocity or pattern speed $\Omega_p$ decreases \citep{Kalnajs1971, Lynden-Bell1972}.

Cosmological simulations provide an opportunity to study how bars form and evolve in galaxies in a cosmological environment, undergoing multiple satellite interactions, flyby events, gas in-fall/outflow, star formation and stellar feedback in the disk, along with a central supermassive black hole and a co-evolving DM halo \citep{Rosas-Guevara.et.al.2021, Rosas-Guevara.et.al.2022,Fragkoudi.et.al.2021, Ansar.et.al.2023b, Fragkoudi2024}. Cosmological simulations can also be used to probe the interaction between the disk and the DM halo at different redshifts which is important for understanding the evolution of the bar -- DM halo interaction.

The properties of the DM halos at redshift $z_r=0$, such as the DM density and DM angular momentum, are the result of the nonlinear evolution of the initial matter density field in the presence of tidal torques \citep{Peebles.1969, Doroshkevich.1970, White.1984, Barnes.and.Efstathiou.1987, Schafer.2009} as well as mergers and interactions with multiple satellite galaxies \citep{Maller.et.al.2002, Vitvitska.et.al.2002, Bett.et.al.2016, Rodriguez-Gomez.et.al.2017}. The DM halo angular momentum is often expressed in a dimensionless form, referred to as the DM halo spin $\lambda$. The halo spin at the virial radius ($r_{vir}$) is given by the \cite{Bullock.et.al.2001} halo spin parameter $\lambda_{B}=L/\sqrt{2}M_{vir} r_{vir} v_{c}(r_{vir})$, where $L$, $M_{vir}$ and $v_{c}(r_{vir})$ is the total halo angular momentum, halo mass and circular velocity at $r_{vir}$. $\lambda_{B}$ is derived from the general expression of halo spin parameter defined by \cite{Peebles.1969}:
\begin{equation} \label{equation:Peebles_spin}
    \lambda = \frac{L |E|^{1/2}}{G M^{5/2}}
\end{equation}
where $L$ is the total angular momentum of the halo, $E$ is the total energy and $M$ is the mass of the DM halo within radius $r$. $\lambda_{P}$ is known as the \citet{Peebles.1969} halo spin when $\lambda$ estimated at the virial radius of the DM halo.

Investigating the properties of halo spin and its connection with galaxy properties has been of interest in many studies \citep{Bett.et.al.2007, Bett.et.al.2010, Bett.Carlos.2012, Bett.et.al.2016}. Halo spin plays an important role determining galaxy sizes, as faster-rotating halos having higher halo spins are associated with larger stellar disks \citep{Kim.Lee.2013, Jiang.et.al.2019,Rodriguez-Gomez.et.al.2022, Perez-Montao.2022, Yang.et.al.2023}. \citet{Romeo.eta.l.2023} studied a large sample of barred S0 galaxies and found a connection between the specific angular momentum of the disk, galaxy morphology and bar structure. \citet{Zjupa.and.Springel.2017} found that the halo spin (both $\lambda_{B}$ and $\lambda_{P}$ at $r_{vir}$) are not much affected by the collapse of the baryons in the DM halos. Recently, \cite{Rosas-Guevara.et.al.2022} using current cosmological hydrodynamical TNG50 simulations \citep{Nelson.et.al.2019, Pillepich.et.al.2019} showed that the halo spin (at $r_{vir}$; $\lambda_{B}$) of barred galaxies is lower than that of the unbarred galaxies at redshift $z_r=0$ (also see \citealt{Izquierdo-Villalba.et.al.2022}). However, there has been no comprehensive study on the origin of the connection between bar strength and halo spin or halo angular momentum. We will see later in Section \ref{sec:bar_role_angular_mom}, that the angular momentum of the stellar disk is more than the angular momentum of the DM halo in the central region of the galaxies (for example, 10 times more inside $r<10$ kpc). Only a significant transfer in angular momentum from the disk to the DM halo can lead to a halo spin change in the halo's inner region. However, no amount of angular momentum transfer can change the halo spin within the virial radius of the halo.

Studies using controlled and isolated galaxy simulations have found different trends of the DM halo properties that aid the formation of bars. \cite{Athanassoula.and.Misiriotis.2002} showed that galaxies with higher DM concentration (having similar stellar to DM ratio) grow stronger, thinner and longer bars due to the transfer of angular momentum from the disk to the DM halo at resonances \citep{Athanassoula.2002}. A central mass concentration in the form of a stellar bulge or a centrally concentrated DM halo can inhibit bar formation \citep{Ostriker.and.Peebles.1973,das.etal.2003, Kataria.and.Das2018, Jang.and.Kim.2023}. Apart from the nature of the DM halo profile, the angular momentum distribution in the central regions of galaxies also plays a major role in bar formation \citep{Long.et.al.2014, Collier.et.al.2018, Collier.et.al.2019a,Collier.et.al.2019b, Collier.Ann-Marie.2021, Kataria2024} and the time of bar buckling \citep{Long.et.al.2014, Collier.et.al.2018,  Kataria.and.Shen.2022, Li.et.al.2023a, Li.et.al.2023b}. 

N-body simulations have shown the formation of a shadow bar or ghost bar in the DM halo during bar formation in the disk \citep{Athanassoula2005, Athanassoula2007, Berentzen.Shlosman.2006, Petersen2016, Collier.et.al.2018, Collier.et.al.2019a, Collier.et.al.2019b, Petersen2019, Petersen2021}. \citet{Li.et.al.2023a} studied the evolution of bars in DM halos by varying halo central densities and halo spins estimated at virial radius. For high spinning halos (at $r_{vir}$), \citet{Li.et.al.2023a} find a significantly large time interval between bar formation and bar buckling, during which the bar strength and pattern speed are constant and the stellar bar aligns with the DM bar and does not transfer angular momentum to the DM halo. However, bar formation is more likely to be affected by the inner halo spin/angular momentum close to the disk rather than at large radii, far from the disk \citep{Kataria.and.Shen.2022, Ansar.et.al.2023}.  

A connection between DM halo spin and the presence of a bar is based on the transfer of angular momentum from the disk to the DM halo through the gravitational interaction with the bar \citep{Debattista.and.Sellwood.2000, Athanassoula.2002, Athanassoula.and.Misiriotis.2002}, thereby increasing the angular momentum of the DM halo. However, the net increase of angular momentum of the DM halo due to the bar--halo interaction is unclear in a cosmologically evolving system, as several other external processes (e.g., satellite mergers, gas accretion) and stellar and DM mass redistribution can equally contribute to the change in angular momentum of the DM halo and the disk. 

In this article, we investigate the connection between the DM halos and stellar bars of disk galaxies in the TNG50 cosmological magneto-hydrodynamical simulation \citep{Nelson.et.al.2019, Pillepich.et.al.2019}. We examine galaxies of different bar strengths (Section \ref{sec:barred_unbarred_sample}). We study the DM halo properties of the sample galaxies at different halo radii and multiple redshifts. We investigate whether the properties of the DM halo and the baryonic disk correlate in our galaxy sample.

We structure this article as follows. We introduce the TNG50 simulations and the galaxy samples in Section \ref{sec:TNG50_galaxy_sample}. We discuss the halo spin of the barred and unbarred galaxies of comparable DM mass at $z_r=0$ in Section \ref{sec:halo_spin_similar_mass}. In Section \ref{sec:bar_role}, we investigate whether the bar plays a role in the increase of DM angular momentum, and in Section \ref{sec:highz_sample23}, we examine the barred galaxies at high redshifts $z_r=0.1$ and 1. We explain issues related to sample selection, biases and convergence in Section \ref{sec:selectionbiasconvergence} and discuss the main findings of this study in Section \ref{sec:discussion}. We present our conclusions in Section \ref{sec:conclusion}. 

\begin{figure*}
\label{fig:galaxy_smaples}
\centering
\includegraphics[width=0.8\textwidth]{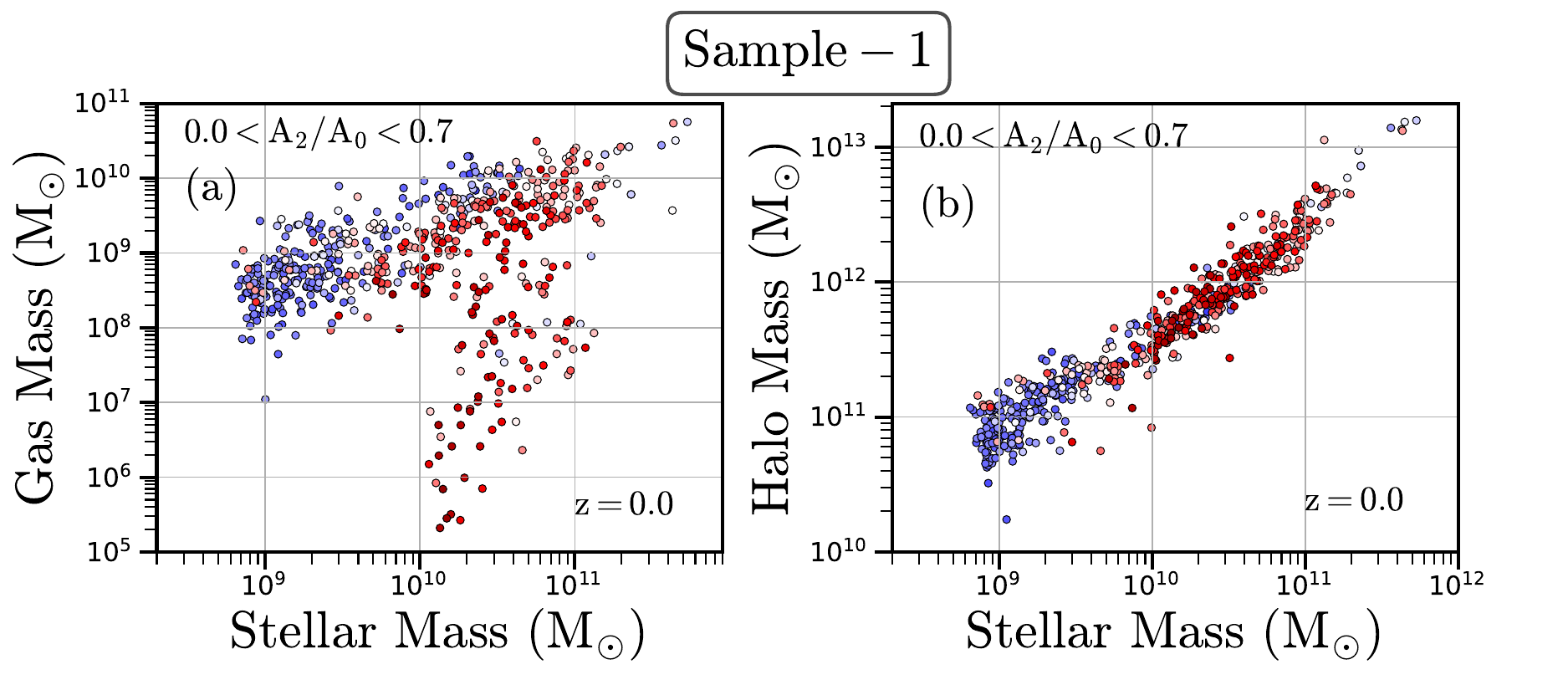} \\
\includegraphics[width=0.8\textwidth]{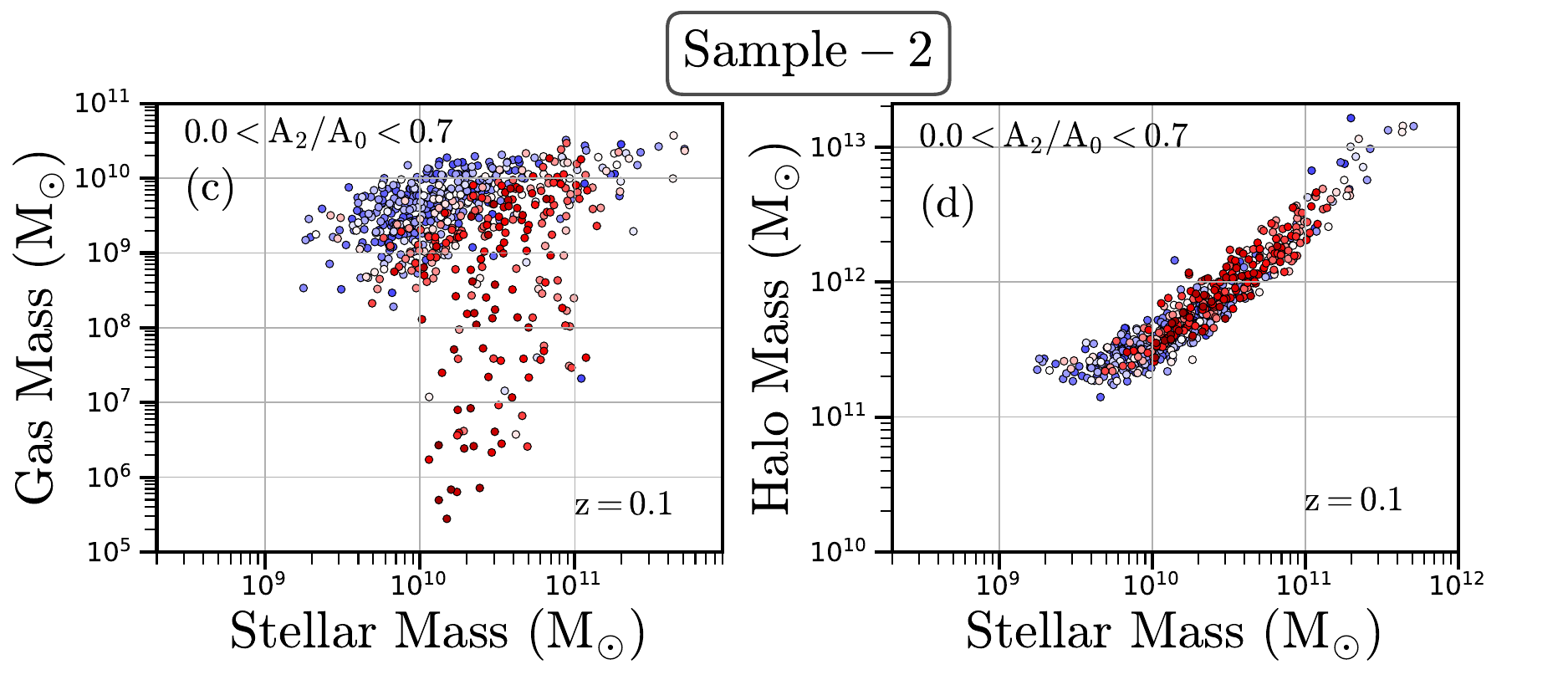}\\
\includegraphics[width=0.8\textwidth]{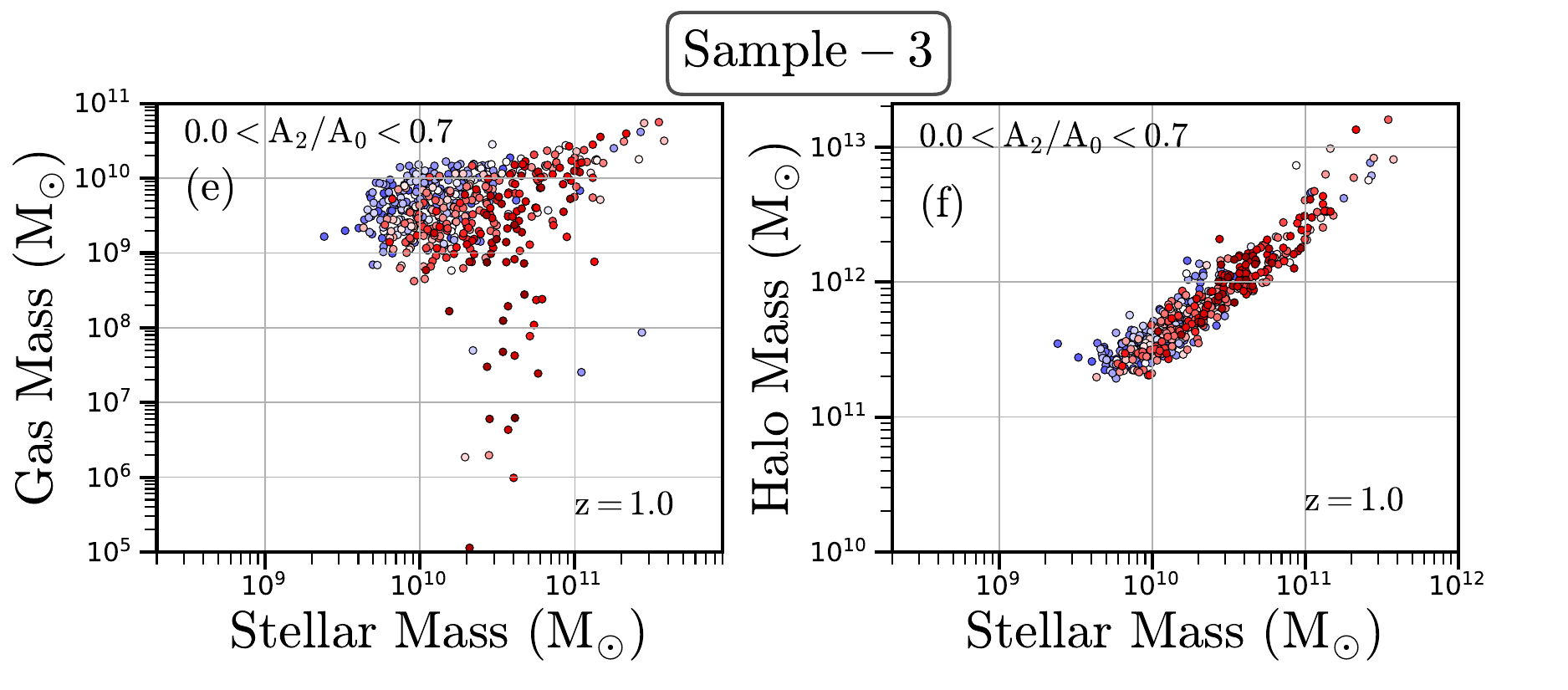} \\
\includegraphics[width=0.3\textwidth]{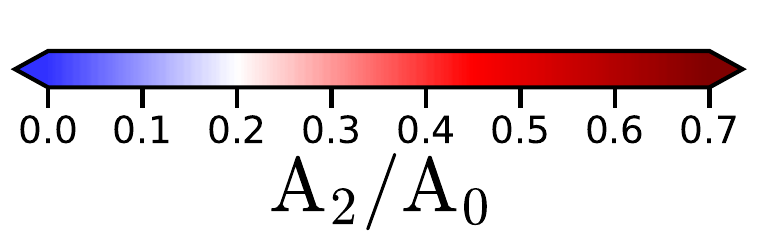}
\caption{{\bf The stellar mass, gas disk mass (panel a, c and e) and dark matter halo mass (panel b, d and f) of the galaxies in our sample at redshift $z_{r}=$0.0 (top row), 0.1 (middle row) and 1.0 (bottom row) with varied bar strengths (in color).} In all panels the bar strength is color-coded with blue for low bar strength $A_2/A_0<0.2$ (see Section \ref{sec:barred_unbarred_sample}), red for high bar strength $A_2/A_0>0.2$ and white for $A_2/A_0=0.2$. In panels a, c and e, a large number of galaxies show low gas content, although most of them have rotationally supported stellar disks. }
\end{figure*}

\section{TNG50 Simulations and galaxy sample} \label{sec:TNG50_galaxy_sample}

We select large samples of galaxies from the cosmological magneto-hydrodynamical simulations IllustrisTNG\footnote{\url{https://www.tng-project.org/}} (The Next Generation) \citep{Weinberger.et.al.2017, Pillepich.et.al.2018} that is run using the moving mesh code \textsf{AREPO} \citep{Springel.2010}.  We use one of the highest resolution data sets TNG50 \citep{Nelson.et.al.2019, Pillepich.et.al.2019}, having a volume of $\sim (51$ $\rm Mpc)^{3}$ and an average mass resolution of $8\times10^{4}$ $\rm M_{\odot}$ for the baryonic particles and $4.5\times10^{5}$ $\rm M_{\odot}$ for the dark matter particles.

We use the pre-computed estimates of dark matter halo mass, stellar disk mass and gas disk mass from the TNG website \citep{Nelson.et.al.2019, Pillepich.et.al.2019}. We adopt the dark matter halo mass from \textsf{SubhaloMassType} which accounts for the total mass of all the particles contained in a Subhalo, and the stellar and gas disk mass using \textsf{SubhaloMassInRadType} parameter, which provides the aggregate of masses of all particles (stars or gas) within twice the stellar half-mass radius. Using the above mass definitions, we form an initial galaxy sample of 1892, 1758 and 1531 galaxies from the TNG50 data set at redshift $z=0, 0.1$ and 1.0, in the galaxy stellar mass range of 6$\times$10$^8 < M_{\star}/M_{\odot}< 10^{12}$ and dark matter halo mass in the range of $10^{10} < M_{DM}/M_{\odot}< 2\times10^{13}$ containing both barred and unbarred galaxies. 

From the above samples, we choose disk galaxies from visual inspection and remove the galaxies that show large deformation in the stellar disk due to ongoing mergers/flyby events of satellites or galaxies that show enhanced off-center concentration of stars in the central region that can generate a high value of the bar strength (used as a probe of a bar; see Section \ref{sec:barred_unbarred_sample}), and will be misinterpreted as a bar.

We avoid galaxies that do not have enough star particles to resolve the central region of the disk. Following the above criteria, we select barred and unbarred disk galaxies from the above samples, with 597 galaxies at redshift $z_{r}=0$, 650 galaxies at $z_{r}=0.1$ and 509 galaxies at $z_{r}=1.0$. Note that we not only track the progenitors of the galaxies at higher redshifts but also construct independent samples of galaxies at the three redshifts.

\subsection{Barred and unbarred galaxies} \label{sec:barred_unbarred_sample}
To find the barred galaxies we first project the disk galaxies in the face-on orientation on the x-y Cartesian coordinate plane such that the maximum angular momentum of the disk is towards the positive z-axis. We rotate the whole system, including the gas and dark matter particles, in the same orientation. Once we have the galaxy disks and the corresponding dark matter halo in the required orientation (with the disk lying in the x-y plane), we use the Fourier decomposition method to identify the triaxial barred structures in the three galaxy samples at different redshifts.

We decompose the face-on stellar surface density $\Sigma_{\star}(r)$ into Fourier modes $\Sigma^{\infty}_{m=0} A_{m}(r) \exp(i m \phi_{m})$. The maximum value of the ratio of the amplitude of the $m=2$ and $m=0$ Fourier mode is the bar strength \citep{Athanassoula.2003}, given as:
\begin{equation}
    \frac{A_{2}}{A_{0}}= \left(\frac{\sqrt{a^{2}_{2} + b^{2}_{2}  }}{\Sigma^{N}_{j=0} m_{\star, j}} \right)_{max}
\end{equation}
where in general for the $m^{th}$ mode, $a_{m}= \Sigma^{N}_{i=1} m_{\star i} \cos(m \theta_{i}) $, $b_{m}= \Sigma^{N}_{i=1} m_{\star i} \sin(m \theta_{i})$, $\theta_{i}$ is the azimuthal coordinate, $m_{\star, j}$ is the mass of the $j^{th}$ star particle, $N$ is the total number of star particles between the radius $r$ to $r+\Delta r$. We adopt $\Delta r=200$ pc. The orientation of the bar is quantified with the phase of the $m=2$ Fourier mode.
\begin{equation}
    \phi_{2}=\tan^{-1}\left( a_2/b_2\right)/2
\end{equation}
One of the ways to find the length of a bar is to find the radius at which the bar phase deviates from a constant value. One of the criteria most frequently used in the literature to differentiate between barred and unbarred galaxies is based on the bar strength: $A_{2}/A_{0} > 0.2$ for barred galaxies and $A_{2}/A_{0} \leq 0.2$ for weakly barred and unbarred galaxies. Following the above criterion, we note that our sample has a slightly larger number of barred galaxies than the number of unbarred galaxies. According to the above definition of bar strength, the bar fraction (fraction of galaxies hosting a bar) is 0.54 at $z_{r}=0$, 0.44 at $z_{r}=0.1$ and 0.54 at $z_{r}=1.0$.

Another important property of a bar is the pattern speed $\Omega_{p}$, the angular velocity of a bar in the disk. We use the recently developed code \textsf{patternSpeed.py} by \citet{Dehnen2023} to measure the pattern speed of bars \citep{Lopez2024, Semczuk2024}. For a detailed discussion on different methods to measure bar length and bar pattern speed see \citet{Athanassoula.and.Misiriotis.2002, Ansar.et.al.2023b}.

In Figure \ref{fig:galaxy_smaples}, we show the stellar mass, gas mass and DM halo mass of the galaxies from the three samples at $z_{r}=0.0, 0.1$ and 1.0, along with their bar strength $A_2/A_0$ in the central region of the disk ($r< 5$ kpc) in color (colorbar range: $0.0<A_{2}/A_{0}<0.7$). Barred galaxies with strength $A_{2}/A_{0}>0.2$ are in shades of red, the unbarred galaxies with strength $A_{2}/A_{0}<0.2$ are in shades of blue and white color is for $A_2/A_0\sim 0.2$.
\begin{deluxetable}{cccc}
\tablenum{1} \label{table:bar_strength_galaxy_no}
\tablecaption{TNG50 galaxies of different bar strength at redshift $z_{r}=0$, 0.1 and 1.0.}
\tablewidth{0pt}
\tablehead{
\colhead{Bar Strength} & \colhead{Sample 1} & \colhead{Sample 2} & \colhead{Sample 3}  \\
\colhead{$A_{2}/A_{0}$ range} & \colhead{($z_{r}=0$)} & \colhead{($z_{r}=0.1$)} & \colhead{($z_{r}=1.0$)}}
\decimalcolnumbers
\startdata
0.0-0.1 & 201 & 224  & 111\\
0.1-0.2 &  72 & 137  & 121\\
0.2-0.3 & 107 & 90  & 68\\
0.3-0.4 & 95  & 78  & 98\\
0.4-0.5 & 74  & 70  & 54\\
0.5-0.6 & 38  & 38  & 41\\
0.6-0.7 & 10  & 13 & 16\\
\enddata
\tablecomments{ Columns: 1. Bar strength $A_{2}/A_{0}$ (see Section \ref{sec:barred_unbarred_sample}); 2. Number of galaxies at redshift $z_{r}=0$; 3. Number of galaxies at redshift $z_{r}=0.1$; 4. Number of galaxies at redshift $z_{r}=1$. A large number of low bar strength galaxies reflects the high number of low mass galaxies where the bar formation is not effective due to reasons like low stellar surface density and high stellar velocity dispersion, shallow DM potential.  }
\end{deluxetable}

We divide the galaxies into 7 groups of different bar strengths between $A_2/A_0=0.0\text{--}0.7$, as in Table \ref{table:bar_strength_galaxy_no}, where we present the number of galaxies in each bar strength interval ($\Delta A_2/A_0=0.1$), for three different redshifts $z_{r}=0.0, 0.1$ and 1.0. The number of barred galaxies decreases for higher bar strengths and the strongest bars ($A_{2}/A_{0}>0.6$) lie in the stellar mass range of $10^{10}\text{--}10^{11}$ M$_{\odot}$ for all three redshifts. On a side note, in Figure \ref{fig:galaxy_smaples} (panel a, c and e), the number of barred and unbarred galaxies within stellar mass $10^{10}<M_{\star}/M_{\odot}<3\times 10^{11}$ and with low gas content, increases for lower redshifts, which distinctly appear as a separate group of points in each panel. We defer the detailed study of the loss of gas content in galaxies to a separate article, the cause of which may be due to multiple processes during evolution, for example, high star formation, ram-pressure stripping of gas from these galaxies, and satellite interactions. We also note a significant number of low mass ($M_{\star}\sim10^9$ M$_{\odot}$) unbarred galaxies (blue) at redshift $z_{r}=0$, which are missing from the higher redshift samples \citep{Flores-Freitas2024}.

\begin{figure*}
\centering
\includegraphics[width=0.8\textwidth]{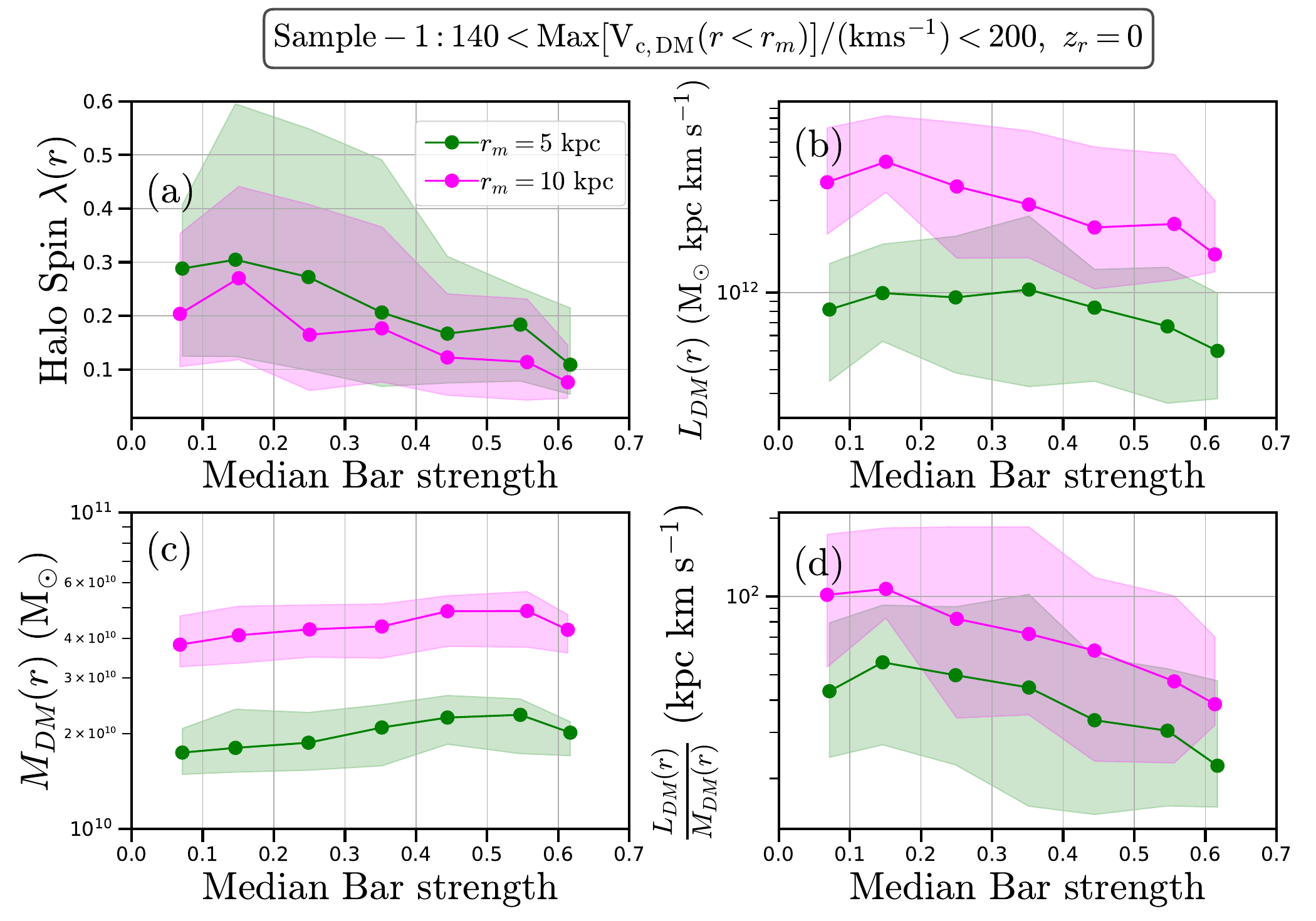}
\caption{{\bf The anti-correlation between bar strength and dark matter halo properties at redshift $z_r=0.0$.} Figure shows the median halo spin (panel a), median DM angular momentum (panel b), median DM mass (panel c) and median specific angular momentum of DM (panel d) for each of the bar strength bins from Table \ref{table:bar_strength_galaxy_no} at redshift $z_{r}=0$, for two radii, $r_{m}=5$ kpc (green, Sample 1r5), 10 kpc (magenta, Sample 1r10), in the velocity range $140< \rm Max\left[V_{c, DM}(r<r_{m})\right]/(km s^{-1})<200$. The solid circles denote the median values and the shaded regions are bounded by the $16^{th}$ and $84^{th}$ percentile curves.}
\label{fig:median_halo_spin_strength_samehalomass}
\end{figure*}

\section{Halo spin of Barred and Unbarred galaxies}\label{sec:halo_spin_similar_mass}
We aim to study the halo spin of barred and unbarred galaxies having similar masses of their DM halos in the central region and close to the baryonic disk. We use the circular velocity ${\rm V_{c, DM}}(r)=\sqrt{G M_{DM}(r)/r}$ of the DM component to select galaxies with similar DM halo mass profiles ($M_{DM}(r)$). We use the selection criteria of $140< \rm Max\left[V_{c, DM}(r<r_{m})\right]/(km s^{-1})<200$ at radii $r_{m}=5$ and 10 kpc, on each of the seven bar strength bins at $z_{r}=0$ from Table \ref{table:bar_strength_galaxy_no}, as it maximises the number of galaxies following this selection criterion. At $z_{r}=0$, we find 25, 21, 42, 34, 29, 29, and 10 galaxies at $r_{m}=10$ kpc (Sample 1r10, hereafter) and 19, 22, 48, 44, 45, 33 and 9 galaxies at $r_{m}=5$ kpc (Sample 1r5, hereafter) for each of the bar strength intervals from Table \ref{table:bar_strength_galaxy_no}. Except for the highest bar strength bin ($0.6<A_2/A_0<0.7$), all other bar strength intervals have a moderate number of galaxies.

\subsection{Galaxies with similar DM circular velocities at redshift $z_{r}=0$}\label{sec:sample1_redshift0}
In Figure \ref{fig:median_halo_spin_strength_samehalomass}, we present the median halo spin $\lambda(r)$ (panel a), median DM angular momentum $L_{DM}(r)=\sqrt{L^{2}_{x}+L^{2}_{y}+L^{2}_{z}}$ (panel b), median DM mass $M_{DM}(r)$ (panel c) and the median DM specific angular momentum $L_{DM}(r)/M_{DM}(r)$ (panel d) using Sample 1r5 (green) and Sample 1r10 (magenta), with their median bar strength in the x-axis of each panel. The solid circles show the median values of each of the above quantities, and shaded regions are bounded by $16^{th}$ and $84^{th}$ percentile curves. Note that, in all of our analysis, we determine the total halo spin at multiple radii using the definition from Equation \ref{equation:Peebles_spin} in the entire article, and not any derived halo spin (for example, $\lambda_{P}$ or $\lambda_{B}$).

In Figure \ref{fig:median_halo_spin_strength_samehalomass} panel c, the median mass of the DM halos for Sample 1r5 and Sample 1r10 is nearly constant over the bar strength intervals (0--0.7), while the median halo spin, the median DM angular momentum and the median DM specific angular momentum (panel b, c and d) decreases as we move from low bar strength to high bar strength galaxies. The median halo spin is higher and the spread in halo spin is larger for the unbarred and weakly barred galaxies ($0<A_2/A_0<0.2$) compared to the strongly barred galaxies ($0.2<A_2/A_0<0.7$). We observe an anti-correlation between bar strength and DM halo spin at redshift $z_r=0$. The overall value of halo spin is lower for larger radii in each of the bar strength intervals, as previously observed in \citet{Ansar.et.al.2023}. The decrease in median halo spin mirrors the decline in median DM angular momentum and DM specific angular momentum for galaxies with stronger bars. Although halo mass slightly increases towards the high bar strength end, this minor mass change is insufficient to account for the decrease in halo angular momentum across the entire range of bar strengths.

For galaxies with higher DM circular velocities, for example, $200< \rm Max\left[V_{c, DM}(r<r_{m})\right]/(km s^{-1})<250$ and at the same radii $r_{m}=5$ and 10 kpc, we observe a similar trend of decrease in median halo spin, DM angular momentum and DM specific angular momentum with higher bar strengths (see Figure \ref{appendix:fig:median_halo_spin_strength_samehalomass2} in Appendix \ref{appendix:sec:decrease_of_halo_spin}). At lower DM circular velocities (e.g., $100< \rm Max\left[V_{c, DM}(r<r_{m})\right]/(km s^{-1})<150$) we do not find a uniform mass distribution of galaxies in this data set, for each of the bar strength intervals, and we cannot compare the halo spin for different bar strength intervals for the lower mass sample (see Figure \ref{appendix:fig:median_halo_spin_strength_samehalomass3} in Appendix \ref{appendix:sec:decrease_of_halo_spin}). The majority of the low-mass unbarred galaxies have low DM halo spin and low median spin values, as seen in Section \ref{sec:halo_spin_all_mass}.

\begin{figure*}
\centering
\includegraphics[width=\textwidth]{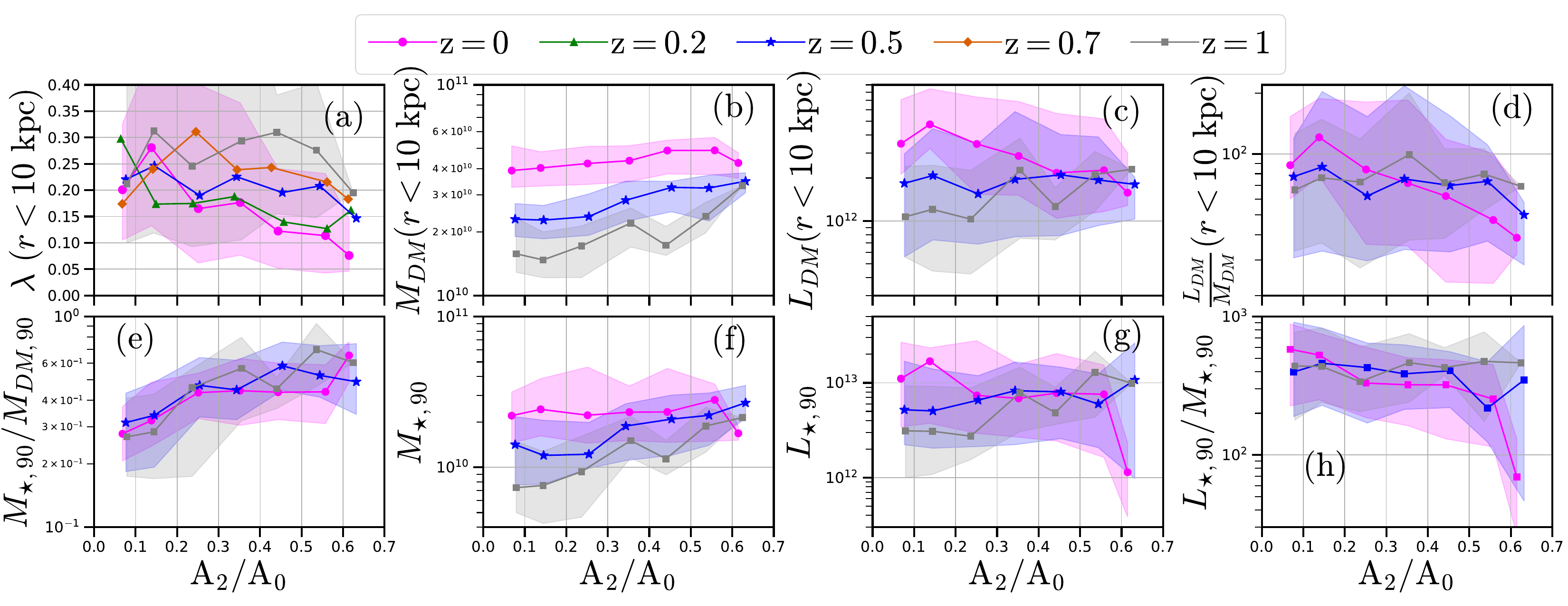}
\caption{{\bf Halo spin of strongly barred galaxies ($A_2/A_0>0.3$) consistently decreases from high to low redshifts, $z_r=1$ to 0.} Figure shows the evolution of halo spin $\lambda$ (panel a), DM halo mass $M_{DM}$ (panel b), DM halo angular momentum $L_{DM}$ (panel c), DM specific angular momentum $L_{DM}/M_{DM}$ (panel d), stellar to DM mass ratio $M_{\star, 90}/M_{DM, 90}$ within $R_{\star, 90}$ (panel e), stellar mass $M_{\star, 90}$ (panel f), stellar angular momentum $L_{\star, 90}$ (panel g) and stellar disk specific angular momentum $L_{\star, 90}/M_{\star, 90}$ (panel h) of the galaxies in Sample 1r10 at $z_r=0$ (magenta), 0.2 (green), 0.5 (blue), 0.7 (orange) and 1 (grey). To avoid crowding, we show three redshifts in all panels except in panel a. }
\label{fig:spin_evolution_highz}
\end{figure*}

\subsection{Tracing the evolution of galaxies at high redshifts}
We track the evolution of the DM halo properties and stellar disk properties of the galaxies in Sample 1r10 (Section \ref{sec:sample1_redshift0}) at four different redshifts, $z_r=0.2$, 0.5, 0.7 and 1.0. 

Figure \ref{fig:spin_evolution_highz} shows the evolution of the DM properties --- halo spin (panel a), mass (panel b), angular momentum (panel c), specific angular momentum (panel d), calculated within $r_m=10$ kpc. We determine the stellar disk properties within $R_{\star, 90}$, the radius at which the stellar mass reaches 90\% of the total mass inside a spherical region (centered at disk center) of radius 30 kpc. These are --- stellar mass $M_{\star, 90}$ (panel f), stellar angular momentum $L_{\star, 90}$ (panel g), stellar specific angular momentum $L_{\star, 90}/M_{\star, 90}$ (panel h) and the ratio of stellar to DM mass $M_{\star, 90}/M_{DM, 90}$ (panel e) as a function of the bar strength $A_2/A_0$ of the galaxies. 

In Figure \ref{fig:spin_evolution_highz} panel a, the median halo spin $\lambda(r<10 \text{ kpc})$ for the strongly barred galaxies decreases steadily from high redshifts to low redshifts. In contrast, the median halo spin of the unbarred galaxies fluctuates around a constant value of $\sim 0.2$ for $0<z_r<1$. In panels b and f, the strongly barred galaxies have higher median DM and stellar mass from an early time compared to the weakly barred and unbarred galaxies that gain mass at a faster rate until $z_r=0$, such that the sample at $z_r=0$ have similar DM mass galaxies across the entire bar strength range ($0<A_2/A_0<0.7$). The DM and stellar angular momentum (panels c and g) are nearly constant for the strongly barred galaxies, while there is a significant gain in DM and stellar angular momentum for the unbarred galaxies from high to low redshifts. The specific angular momentum of the DM and the stars are similar for the barred and unbarred galaxies at the high redshifts ($z_r=1$). However, at lower redshifts (e.g., $z_r=0$), there is a gradient in the specific angular momentum, where the barred galaxies have lower DM and stellar specific angular momentum than the unbarred ones. Finally, in panel e, the median of the stellar to DM mass ratio increases from low to high bar strengths, and the ratio is nearly constant across the entire redshift range. To avoid crowding, we show three redshifts in all the panels (except panel a). 

The anti-correlation between halo spin and bar strength of galaxies is most prominent at $z_r=0$. The anti-correlation becomes weaker with increase in redshifts till $z_r=1$, where the median halo spin is nearly similar for all bar strengths. 

\subsection{Origin of the bar strength - halo spin anti-correlation}\label{sec:origin_halo_spin}
Not all galaxies in Sample 1r10 with strong bars at $z_r=0$ host strong bars at high redshifts. In some galaxies, bars form much later than $z_r=1$ and gain strength towards the end of their evolution. As a result, the number of galaxies in different bar strength intervals evolves with time. To avoid the change in the number of galaxies across different bar strength bins, we study two distinct classes of galaxies from Sample 1r10 in more detail --- 1. 10 strongly barred galaxies with $A_2/A_0>0.6$ at $z_r=0$ and 2. 13 unbarred/weakly barred galaxies with $A_2/A_0<0.2$ at $z_r=0$ and during most of the time of their evolution. Using the TNG50 data we can calculate the halo spin at specific redshifts as the potential of the DM particles are available only at certain redshifts (for example, $z_r=0$, 0.1, 0.2, 0.3, 0.4, 0.5, 0.7, 1.0, 1.5, 2.0).
\begin{figure}
\centering
\includegraphics[width=\columnwidth]{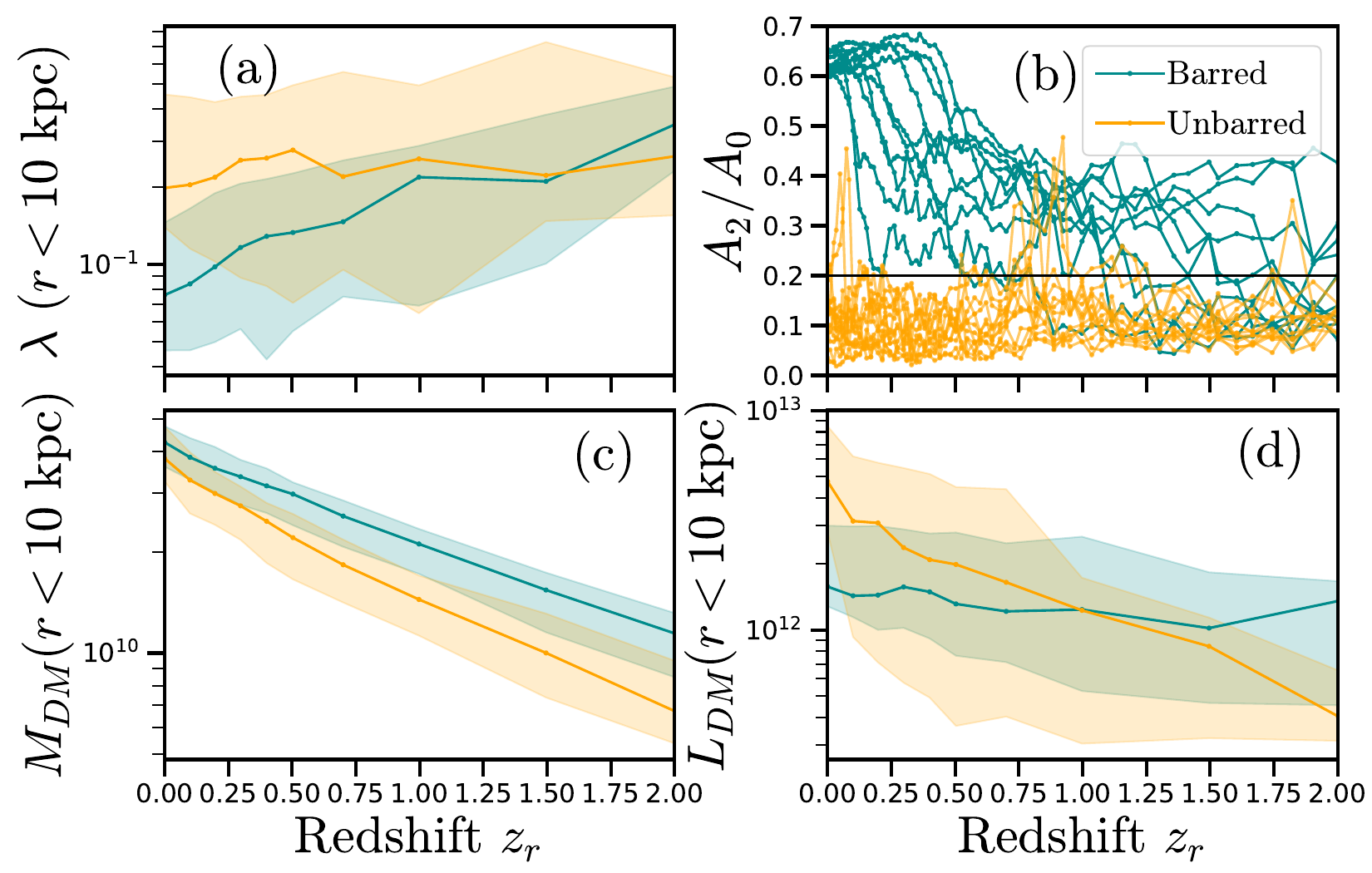}
\caption{{\bf Halo spin in DM halos hosting barred galaxies (green) decreases faster with redshift than DM halos hosting unbarred galaxies (yellow).} The Figure shows the evolution of median halo spin (panel a), bar strength (panel b), DM mass (panel c) and DM angular momentum (panel d), all estimated within $r<10$ kpc. The higher DM mass of barred galaxies leads to lower halo spin, even though the angular momentum of unbarred galaxies is higher than the barred ones. }
\label{fig:spin_angularmom_mass_strength_evolution_highz}
\end{figure}

In Figure \ref{fig:spin_angularmom_mass_strength_evolution_highz}, we show the evolution of median halo spin (panel a), median DM mass (panel c), median DM angular momentum (panel d) along with the evolution of their bar strengths (panel b) with redshift $z_r$. The strongly barred galaxies are green, and the unbarred galaxies are golden. As seen from panel b, the strongly barred galaxies cross $A_2/A_0>0.2$ at some point during their evolution and remain strong with $A_2/A_0>0.6$ at $z_r=0$. The unbarred and strongly barred galaxies have similar median halo spins at high redshifts ($z_r>1$). However, towards low redshifts, the median halo spin of the strongly barred galaxies decreases significantly lower than the unbarred galaxies. The median spin of the unbarred galaxies decreases at a very slow rate and can be considered nearly constant. The distinct values of median halo spin for the strongly barred and unbarred galaxies result from the different rates of rise in DM angular momentum and DM mass within $r_m=10$ kpc. Even though the median DM mass for the two samples is similar at $z_r=0$, the unbarred galaxies have distinctly lower DM mass at high redshifts. Additionally, the rate of increase in DM angular momentum for the strongly barred galaxies is relatively slower than the unbarred galaxies (see panel d). 

In conclusion, barred and unbarred galaxies that have similar DM mass at $z_r=0$, show a divergence in DM mass at high redshifts. The massive galaxies tend to host stellar bars while the less massive galaxies do not. The rise in DM mass within a fixed radius ($r_m=10$ kpc) offsets the gradual rise in DM angular momentum, leading to a more rapid decrease in the median spin of barred galaxies compared to unbarred ones. Even though the DM mass in unbarred galaxies increases more rapidly than the unbarred galaxies, the median DM mass is lower (for $z_r>0$) for the unbarred galaxies. Additionally, the median DM angular momentum in unbarred galaxies grows faster (after $z_r<1$), resulting in higher median halo spin for unbarred galaxies relative to the barred ones.

\section{Role of bar in the increase in DM angular momentum}\label{sec:bar_role}
We investigate the transfer of angular momentum between the stellar bar and their host DM halos using the 10 strongly barred galaxies in Sample 1r10 with $A_2/A_0>0.6$ at $z_r=0$. These are the same barred galaxies used in Figure \ref{fig:spin_angularmom_mass_strength_evolution_highz}. 

\subsection{Evolution of disk and DM angular momentum}\label{sec:bar_role_angular_mom}
As the bar strength in each galaxy evolves with redshift, we track the evolution of bar pattern speed $\Omega_p$, stellar and DM angular momentum, and the evolution of the stellar and DM mass within a spherical region of radius 10 kpc centered at the disk center of mass. As the bars appear in the disks at different times, we need to normalise the redshift scale with the redshift of bar formation $z_{form}$. We define $z_{form}$ as the redshift after which the bar strength crosses $A_2/A_0>0.2$ and the bar phase $\phi_2$ is nearly constant within the bar length.  

\begin{figure}
\centering
\includegraphics[width=0.9\columnwidth]{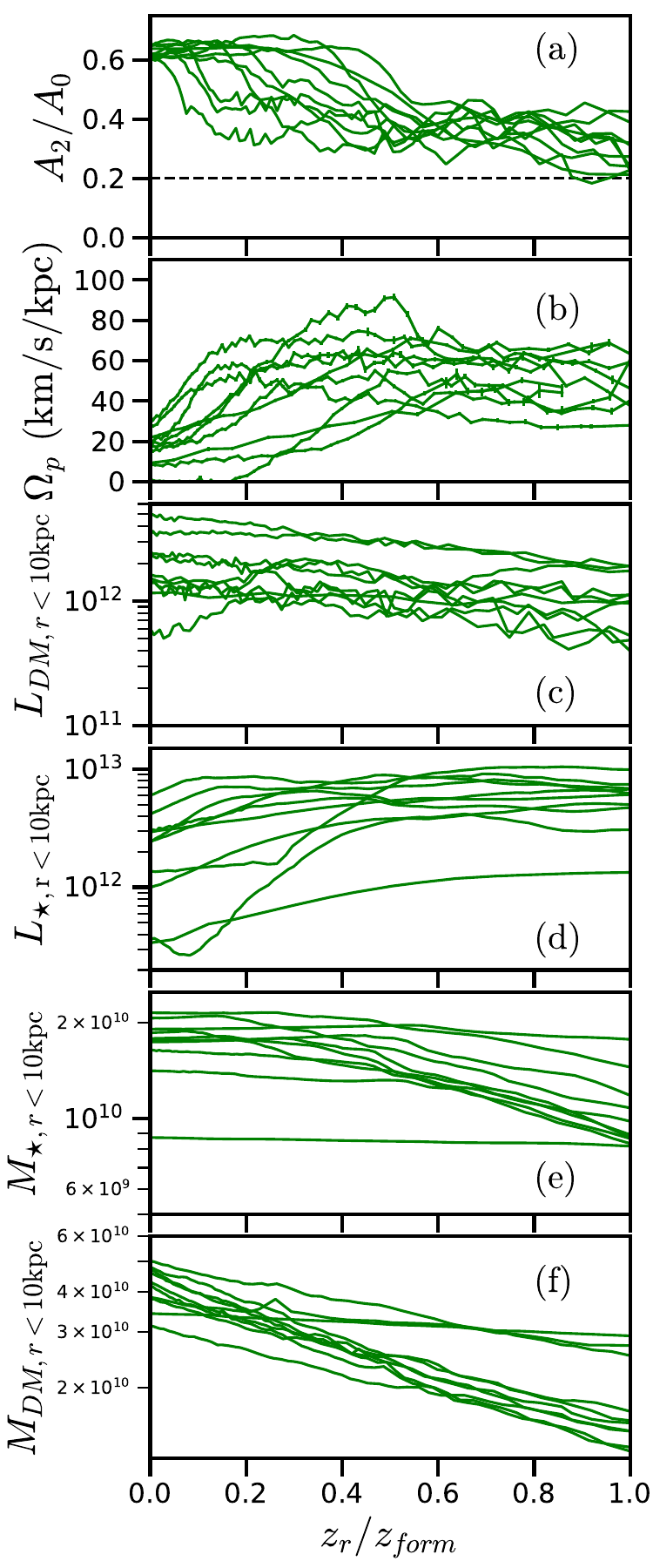}
\caption{{\bf Decrease in bar pattern speed $\Omega_p$ (panel b) and stellar disk angular momentum $L_{\star}$ (panel d) with rise in bar strength $A_2/A_0$ (panel a) and DM angular momentum $L_{DM}$ (panel c).} The Figure also shows the simultaneous rise in stellar $M_{\star}$ and DM mass $M_{DM}$ within a radius of 10 kpc. The redshift is normalised with the bar formation redshift $z_{form}$. }
\label{fig:strength_evolution_highz}
\end{figure}

Figure \ref{fig:strength_evolution_highz} shows the evolution of bar strength in the 10 strongly barred galaxies (panel a), along with the evolution bar pattern speed (panel b), DM angular momentum (panel c), stellar angular momentum (panel d), stellar mass (panel e) and DM mass (panel f) within $r_m<10$ kpc. As the bar strength increases, the stellar angular momentum decreases even though the stellar mass in the central region increases (see panel e). Often a rise in bar strength is accompanied by an increase in DM angular momentum. This may be because the bar can transfer angular momentum from the disk to the DM halo through tidal torques on the DM particles close to the central region, thereby slowing down its pattern speed in the process. Furthermore, the rise in DM angular momentum may also be due to the rise in the DM mass (see panel f). The decrease in the bar pattern speed and the decrease in stellar angular momentum with the rise in bar strength suggests that the bar is transferring angular momentum to the DM halo. 

\subsection{Correlations between bar strength and stellar and DM angular momentum}
We quantify the relationships between the rise in DM angular momentum, DM mass and stellar mass and the decrease in stellar angular momentum, alongside the strengthening of the bar, using the Pearson Correlation Coefficient $R$ in our study. We use the 10 strongly barred galaxies and the 13 unbarred galaxies mentioned in Section \ref{sec:sample1_redshift0} and \ref{sec:origin_halo_spin}. We estimate five correlation coefficients for all the galaxies in both samples over a specific redshift interval. We evaluate the correlation $R$ between 1. $L_{\star}$ and $A_2/A_0$, 2. $L_{DM}$ and $L_{\star}$, 3. $L_{\star}$ and $M_{\star}$, 4. $L_{DM}$ and $M_{DM}$ and 5. $L_{DM}$ and $A_2/A_0$, where all quantities are calculated within $r_m=10$ kpc for a specific redshift interval. For the barred galaxies, we choose the redshift interval that satisfies the condition: $A_2/A_0>0.35$ and $z_r>1$, and for the unbarred galaxies, we choose the redshifts: $z_r>1$. The above selection of intervals accounts for the galaxies' strongly barred phase, where the angular momentum transfer from the disk to the DM halo is probable and keeps a similar upper limit of redshift for both barred and unbarred systems. 

\begin{figure}
\centering
\includegraphics[width=\columnwidth]{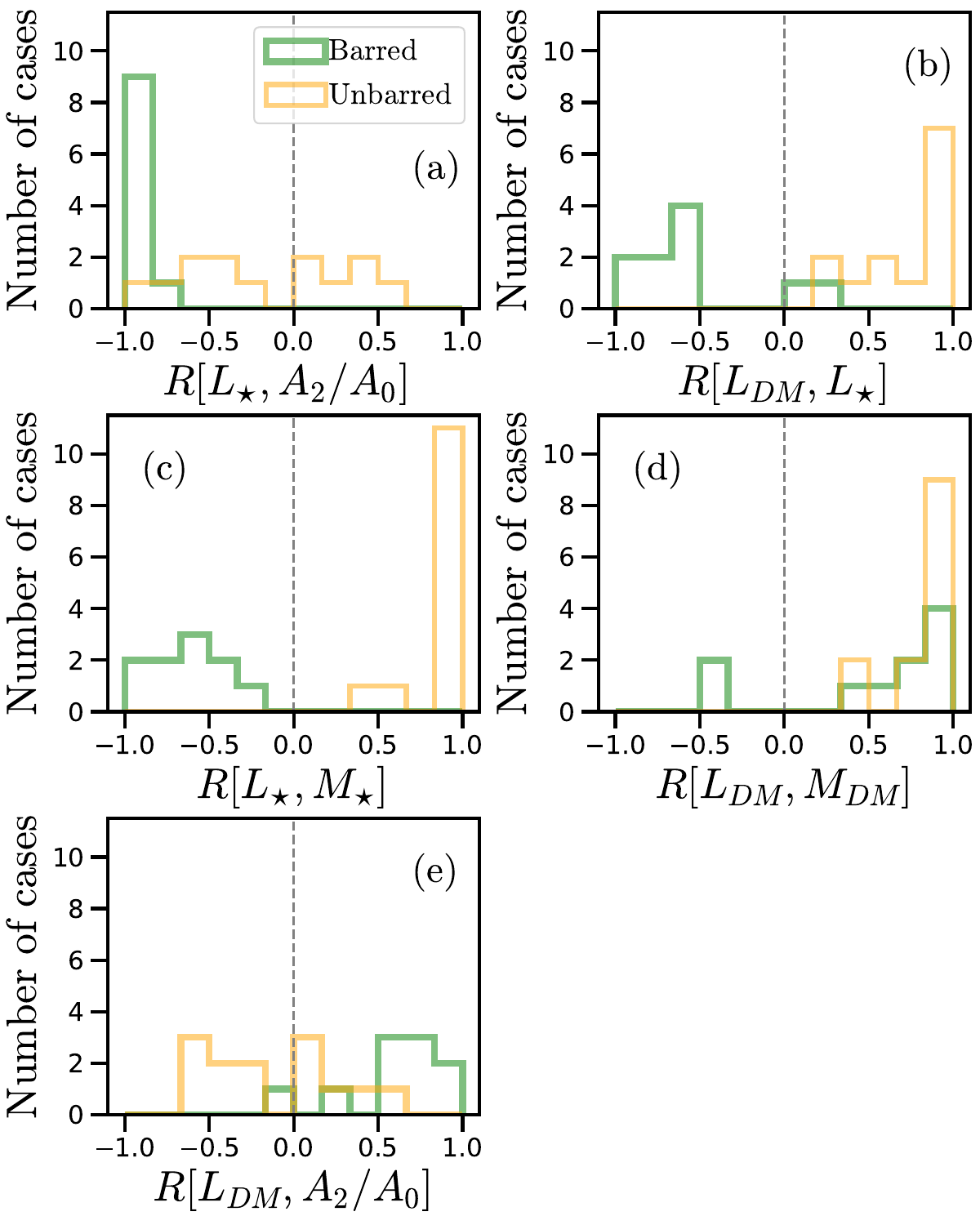}
\caption{{\bf Increase in bar strength leads to the decrease in disk angular momentum and coincides with the rise in DM angular momentum.} {\bf X-axis of all panels:} (a) Pearson correlation coefficient $R$ for stellar angular momentum $L_{\star}$ and bar strength $A_2/A_0$; (b) DM angular momentum $L_{DM}$ and $L_{\star}$; (c) $L_{\star}$ and stellar mass $M_{\star}$; (d) $L_{DM}$ and DM mass $M_{DM}$ and (e) $L_{DM}$ and $A_2/A_0$. All quantities are estimated within $r=10$ kpc. The y-axes show the number of systems for barred (green) and unbarred galaxies (yellow).}
\label{fig:correlations}
\end{figure}
In Figure \ref{fig:correlations}, we present the histograms of the above correlation coefficients in different panels for barred galaxies (green) and unbarred galaxies (yellow). Panel a shows that the rise in bar strength and the decrease in stellar angular momentum for barred galaxies leads to $R\sim -1$, while for the unbarred galaxies, $R$ is spread over a broader range with no significant correlation. The correlation between $L_{DM}$ and $L_{\star}$ (panel b), and $L_{\star}$ and $M_{\star}$ (panel c) show negative values of $R$ for barred galaxies and positive value of $R$ for unbarred galaxies. For both barred and unbarred galaxies, the correlation $R$ between $L_{DM}$ and $M_{DM}$ is mostly positive, except for 2 exceptional cases of barred systems. This positive correlation shows that in both barred and unbarred systems, there is an increase in DM angular momentum and mass with time. However, for barred galaxies, the rise in bar strength is mostly positively correlated with a rise in $L_{DM}$, while no correlation is seen for unbarred galaxies (panel e). 

\begin{figure*}
\centering
\includegraphics[width=0.8\textwidth]{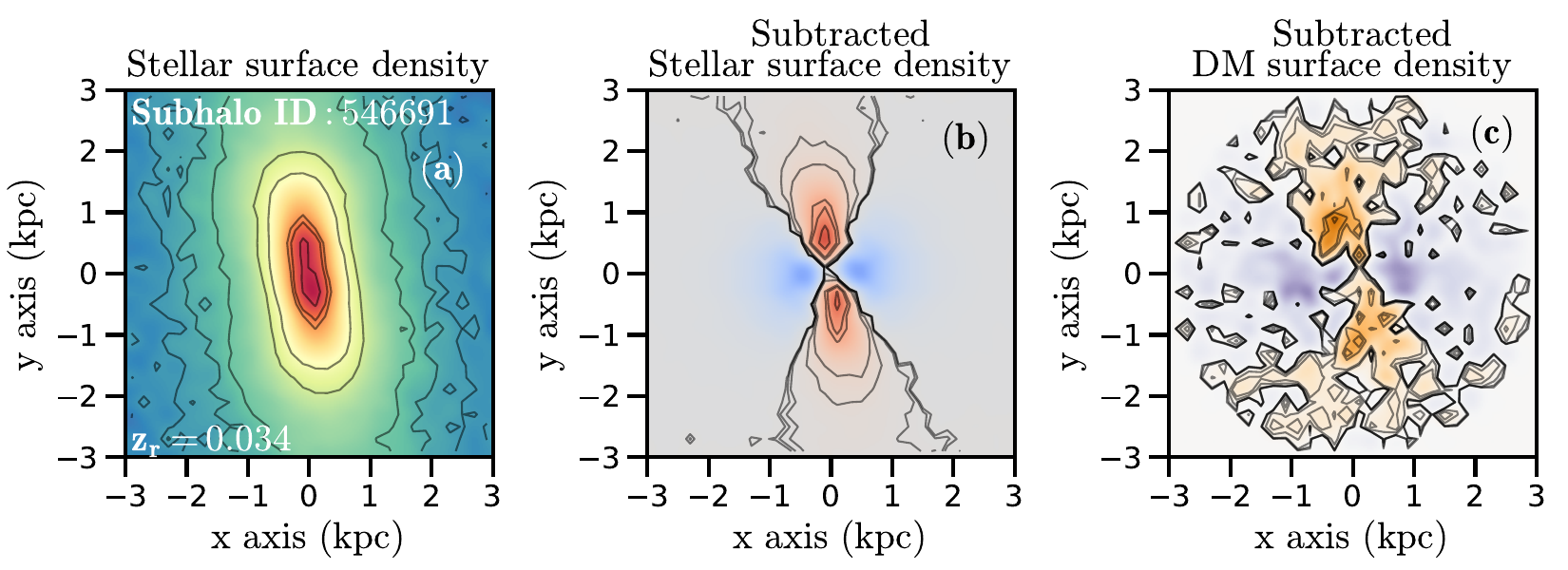}\\
\vspace{-0.1cm} \hspace{1.3cm}
\includegraphics[width=0.2\textwidth]{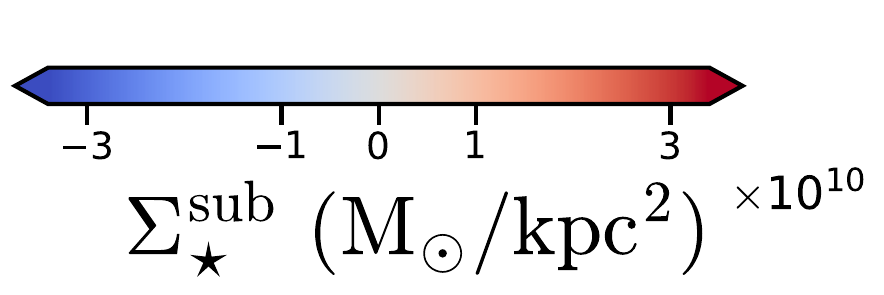}\\
\vspace{-0.1cm}
\includegraphics[width=0.8\textwidth]{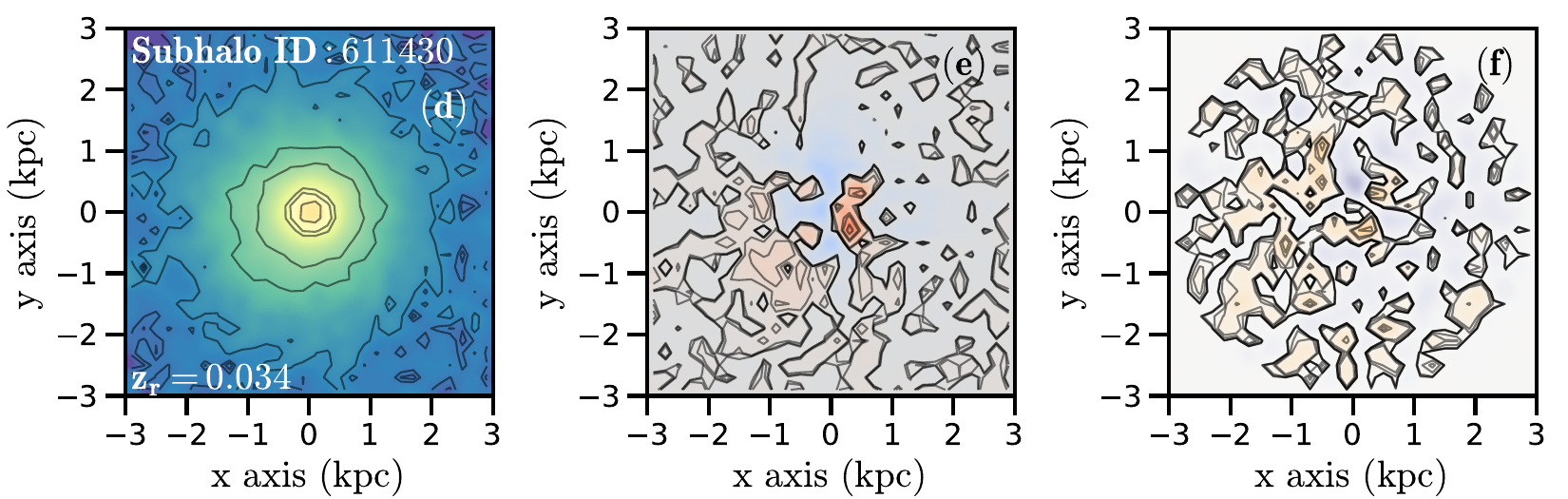}\\
\vspace{-0.1cm} \hspace{1.3cm}
\includegraphics[width=0.2\textwidth]{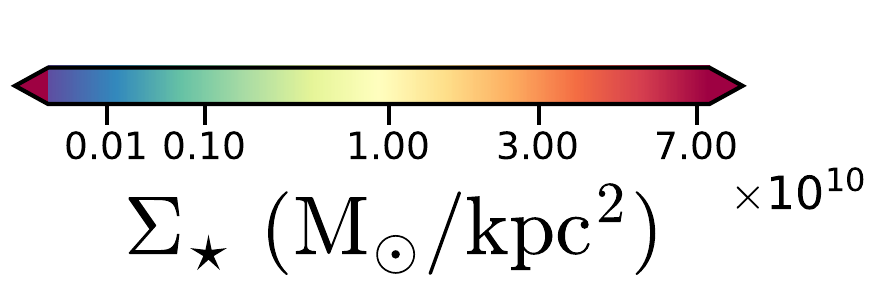} \hspace{0.95cm}
\includegraphics[width=0.2\textwidth]{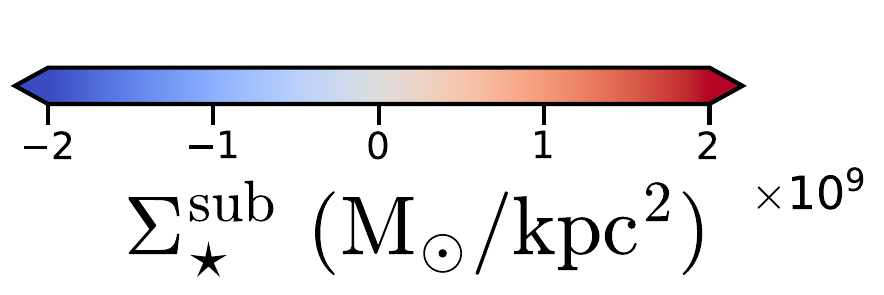} \hspace{0.95cm}
\includegraphics[width=0.2\textwidth]{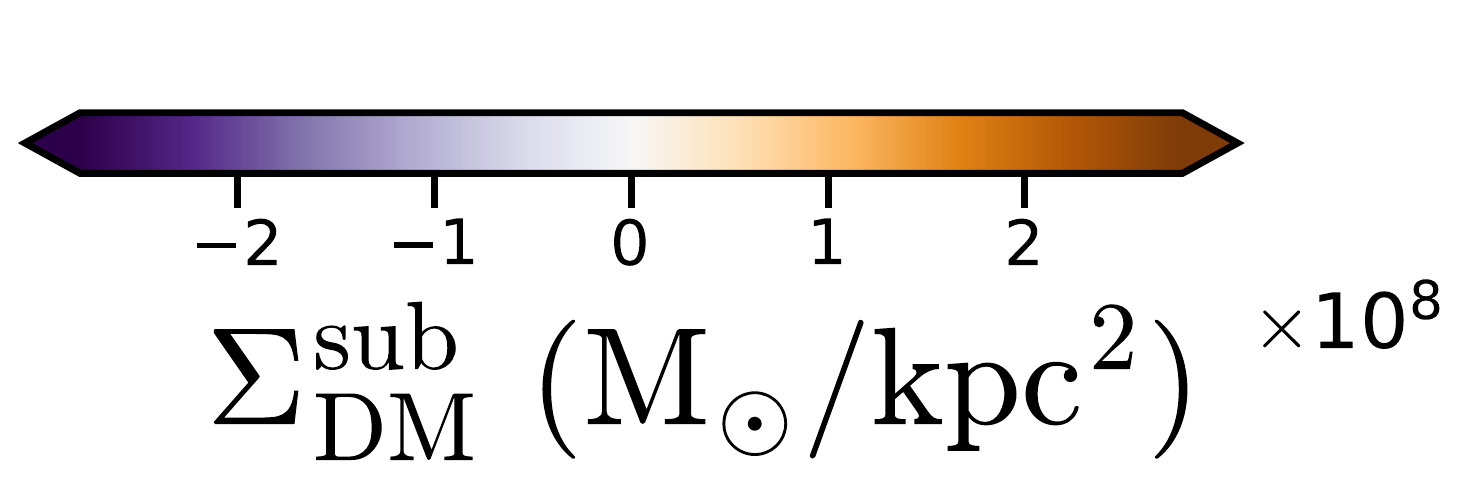}
\caption{{\bf Shadow DM bars are aligned with the stellar bars in strongly barred galaxies. Top panels:} (a) face-on stellar surface density of a strongly barred galaxy with Subhalo ID 546691, $z_r=0.034$; (b) subtracted stellar surface density $\Sigma_{\star}^{sub}(r)$ map showing positive (red) and negative (blue) regions; (c) subtracted DM surface density $\Sigma_{DM}^{sub}(r)$ map showing positive (brown) and negative (purple) regions. {\bf Bottom panels:} same as the top panel for an unbarred galaxy with Subhalo ID 611430. Colorbars are common for panels a and d, and among panels c and f. The quadruple moment in $\Sigma_{\star}^{sub}(r)$ map (panel b) for a strongly barred galaxy is also traced by the DM distribution $\Sigma_{DM}^{sub}(r)$ (panel c), however, no quadrupole moment is seen for the unbarred galaxy. }
\label{fig:DM_bars}
\end{figure*}
\subsection{Shadow Dark Matter Bar}\label{sec:dark_bar}
More direct evidence of the interaction between the stellar bar and the surrounding DM halo is seen through an over-density in DM along the bar's major axis. 

After aligning the disk angular momentum towards the $z$-direction, we determine the surface density maps in a 6 kpc$\times$6 kpc region centered at the disk center, with thickness $|z|<0.5$ kpc. Figure \ref{fig:DM_bars} shows the stellar surface density ($\Sigma_{\star}(r)$) map of a strongly barred galaxy with Subhalo ID: 546691 at $z_r=0.034$ (panel a) and an unbarred galaxy with Subhalo ID: 611430 at $z_r=0.034$ (panel d), with their common colorbar. Next, we create the subtracted stellar density $\Sigma_{\star}^{sub}(r)$ maps for the two galaxies, that is generated by subtracting the mean stellar density $\bar{\Sigma}_{\star}(r)$ over concentric annular rings (centered at the disk center) of width $\Delta r=0.2$ kpc and thickness $|z|<0.5$ kpc, from the total surface density:  $\Sigma_{\star}^{sub}(r)=\Sigma_{\star}(r)- \bar{\Sigma}_{\star}(r)$. Panels b and e show the subtracted stellar surface density $\Sigma_{\star}^{sub}(r)$ maps for the barred and unbarred galaxies, along with their colorbars. Similarly, we generate the corresponding subtracted DM density $\Sigma_{DM}^{sub}(r)$ maps and present them in panels c and f, with their common colorbar. 

In Figure \ref{fig:DM_bars} panel b, the two-lobed structure in $\Sigma_{\star}^{sub}(r)$ map is due to the asymmetric structure of the stellar bar in the $\Sigma_{\star}(r)$ map (panel a). In panel c, we see a similar over-density in the underlying DM $\Sigma_{DM}^{sub}(r)$ map that is aligned with the stellar bar. This two-lobed structure indicates a bar in the DM density distribution, and we have checked that it is present for all the strongly barred galaxies in Sample 1r10 with $A_2/A_0>0.6$ at $z_r=0$. We follow the evolution of the two-lobed structure in the DM distribution and find that it starts after the bar formation time. For unbarred galaxies, such a structure is not observed (for example, see panel f). The $\Sigma_{DM}^{sub}(r)$ maps are noisy due to the resolution of the simulation. \citet{Ash2024} recently characterized the DM bar strength for the barred galaxy sample from \citet{Rosas-Guevara.et.al.2022}, and found that the pattern speed of the stellar and DM bar in a galaxy in Subbox-0 (a sub-volume in TNG50 with a box size of 7.5 $h^{-1}$ Mpc) is nearly synchronous, and the DM bar is aligned with the stellar bar during 8 Gyrs of evolution. Using high-resolution zoom-in simulations, we will be able to study DM bars with a level of accuracy comparable to that achieved in high-resolution N-body simulations (for example, see Figure 7 in \citealt{Collier.et.al.2019a}) in the future.

\begin{figure*}
\centering
\includegraphics[width=0.75\textwidth]{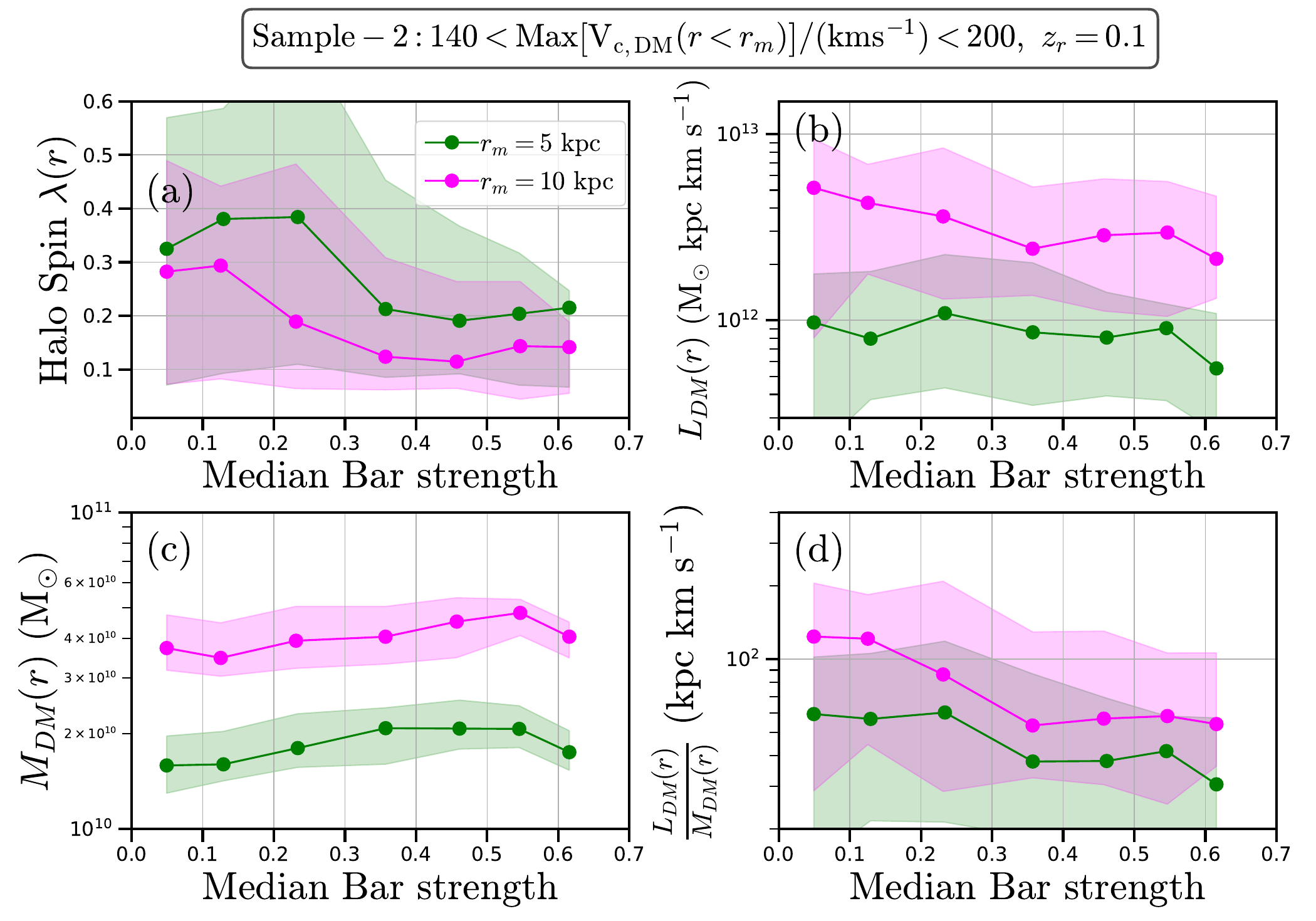} \\
\includegraphics[width=0.75\textwidth]{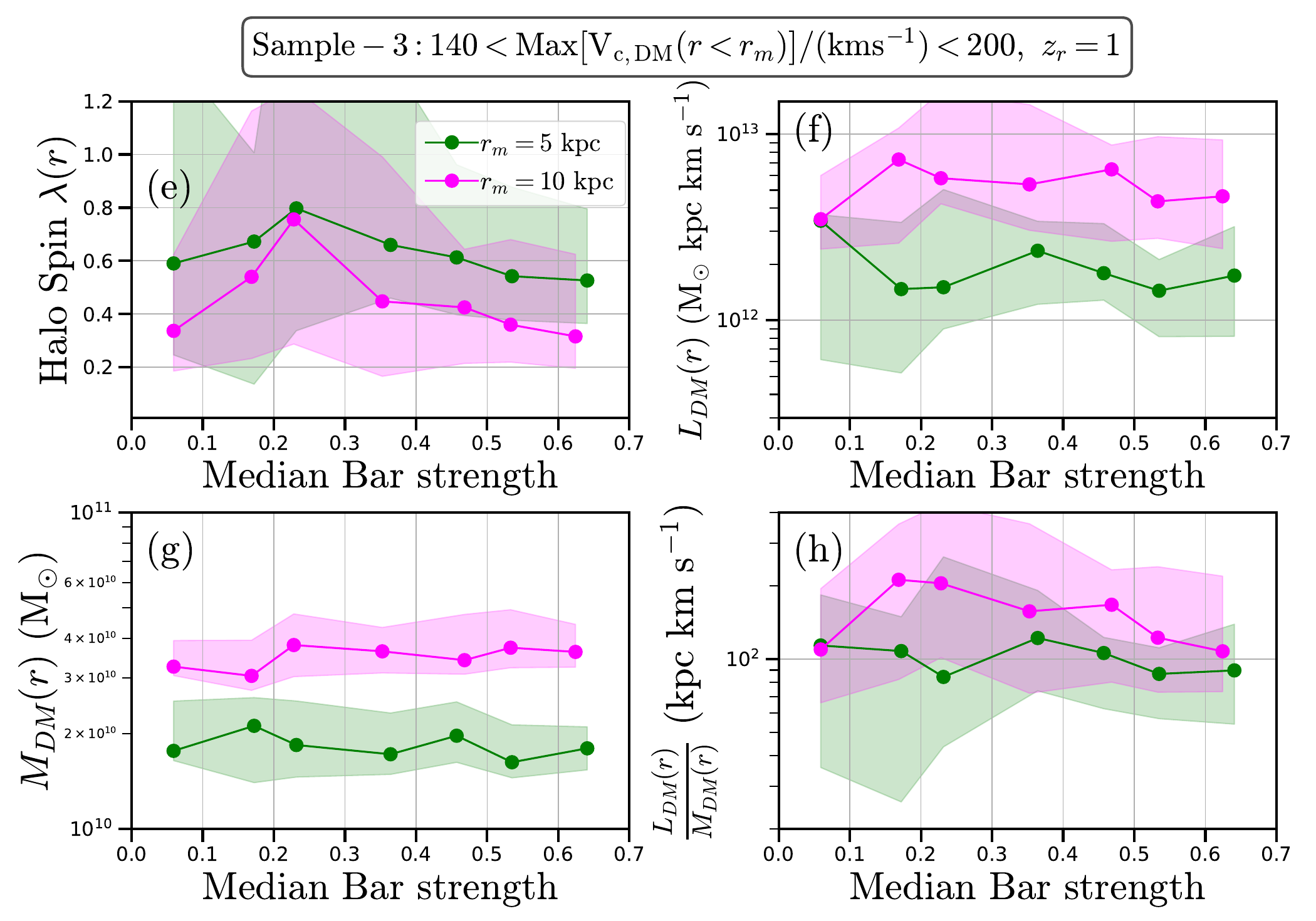}
\caption{{\bf The connection between bar strength and dark matter halo properties at high redshift $z_r=$0.1 (panel a-d) and 1.0 (panel e-h).} Same as Figure \ref{fig:median_halo_spin_strength_samehalomass}, for the velocity range: $140< \rm Max\left[V_{c, DM}(r<r_{m})\right]/(km s^{-1})<200$ at $z_r=0.1$ and 1.0. At $z_r=1.0$, the bar strength -- halo spin connection is more complex.}
\label{fig:median_halo_spin_strength_samehalomass_highz}
\end{figure*}

\section{Halo spin of galaxies in Sample-2 and Sample-3}\label{sec:highz_sample23}
We investigate if there is an anti-correlation between the halo spin and the bar strength for galaxies with comparable DM masses at high redshifts, similar to $z_{r}=0$ (Section \ref{sec:barred_unbarred_sample}). Here we examine the barred and unbarred galaxies at higher redshifts $z_{r}=0.1$ and 1.0 by forming samples of galaxies having similar DM mass using Sample 2 and Sample 3 from Table \ref{table:bar_strength_galaxy_no}. We apply the same selection criteria from Section \ref{sec:sample1_redshift0} based on the DM circular velocities: $140< \rm Max\left[V_{c, DM}(r<r_{m})\right]/(km s^{-1})<200$, on each of the seven bar strength bins in Sample 2 at $z_{r}=0.1$ and Sample 3 at $z_r=1$. We find 37, 58, 46, 39, 40, 28 and 12 galaxies at $r_{m}=10$ kpc (Sample 2r10 hereafter) and 37, 40, 42, 42, 49, 34, 12, galaxies at $r_{m}=5$ kpc (Sample 2r5 hereafter). At $z_{r}=1.0$, we find lesser number of galaxies in Sample-3 in the above circular velocity range: 5, 7, 11, 19, 23, 28 and 11 galaxies for $r_m=5$ kpc (Sample 3r5 hereafter) and 8, 19, 15, 24, 33, 38 and 15 galaxies for $r_m=10$ kpc (Sample 3r10 hereafter). We present the median halo spin, median DM angular momentum, median DM mass and median DM specific angular momentum for the seven bar strength bins in Figure \ref{fig:median_halo_spin_strength_samehalomass_highz}, with Samples 2r5 (panels a--d, green), 2r10 (panels a--d, magenta), 3r5 (panels e--h, green) and 3r10 (panels e--h, magenta), (similar to Figure \ref{fig:median_halo_spin_strength_samehalomass}). We have also conducted tests with a slightly different circular velocity selection criterion of $130< \rm Max\left[V_{c, DM}(r<r_{m})\right]/(km s^{-1})<180$ with larger number of galaxies at $z_r =1$ to find 18, 24, 16, 39, 33, 32 and 14 galaxies for $r_{m}=10$ kpc and 8, 12, 12, 24, 30, 35 and 15 galaxies for $r_{m}=5$ kpc for the seven bar strength bins and found similar trends in the results. 

Figure \ref{fig:median_halo_spin_strength_samehalomass_highz} panel a shows that at $z_r=0.1$, the halo spin and bar strength follow a similar anti-correlation as in the case of $z_r=0$. However, the median halo spin in each bar strength bin at $z_r=0.1$ is higher than $z_r=0$. In the bar strength range $0.6<A_2/A_0<0.7$, the median halo spin in Sample 2r5 (Figure \ref{fig:median_halo_spin_strength_samehalomass_highz} panel a) is twice the median spin in Sample 1r5 (Figure \ref{fig:median_halo_spin_strength_samehalomass} panel a). For $0.1<A_2/A_0<0.2$, the median halo spin for Sample 2r5 is 4/3rd times the median spin for Sample 1r5. Similarly, even higher halo spins are seen for higher redshifts ($z_r=1$, panel e). The median DM mass is nearly constant across the bar strength range 0--0.7 (panel c), while the median DM angular momentum is nearly constant for $r_m=5$ kpc, but has a decreasing trend for $r_m=10$ kpc (panel b). The median specific angular momentum of the DM halo for both radii show similar decreasing trends with higher bar strengths (panel d). At $z_r=1$ (panel e), the halo spin for the high bar strength galaxies ($0.3<A_2/A_0<0.7$) and very low bar strength galaxies ($A_2/A_0<0.1$) is nearly similar, however in the transition region of low to high bar strength galaxies ($0.1<A_2/A_0<0.3$) the halo spin increases to high values. Even though there is a significant difference between the high-spinning weakly-barred galaxies ($0.1<A_2/A_0<0.3$) and low-spinning barred galaxies ($A_2/A_0>0.3$), the anti-correlation is not followed by galaxies with $A_2/A_0<0.2$.

To summarise Section \ref{sec:highz_sample23}, the anti-correlation between the median halo spin and median bar strength exists at low redshifts ($z_r=0.0$ and 0.1), but a clear anti-correlation is not there for high redshift $z_r=1.0$ galaxies. The relation between median halo spin and bar strength is more complex at $z_r=1.0$. The high redshift galaxies have higher median halo spin values than the low redshifts for nearly all bar strength ranges. The increase in halo spin at high redshifts is also evident from Section \ref{sec:origin_halo_spin}, where the lower DM mass at high redshifts (compared to low redshifts) contributes to the higher halo spin. This indicates that the galaxies in which bars formed and sustained till low redshifts have notable differences in their stellar disk mass, DM halo mass and DM angular momentum during evolution.

\section{Sample selection, Biases and Convergence}  \label{sec:selectionbiasconvergence}

\subsection{Sample selection and biases}\label{sec:halo_spin_all_mass}
In this Section, we present the reason for a sample selection criterion based on DM mass and point out the biases that can lead to unclear correlations in DM halo spin and bar strengths. We relax the criterion based on circular velocities of DM used in Section \ref{sec:halo_spin_similar_mass} and consider all the barred and unbarred galaxies from the three samples in Table \ref{table:bar_strength_galaxy_no} irrespective of their DM halo mass.  

\begin{figure*}
\centering
\includegraphics[width=\textwidth]{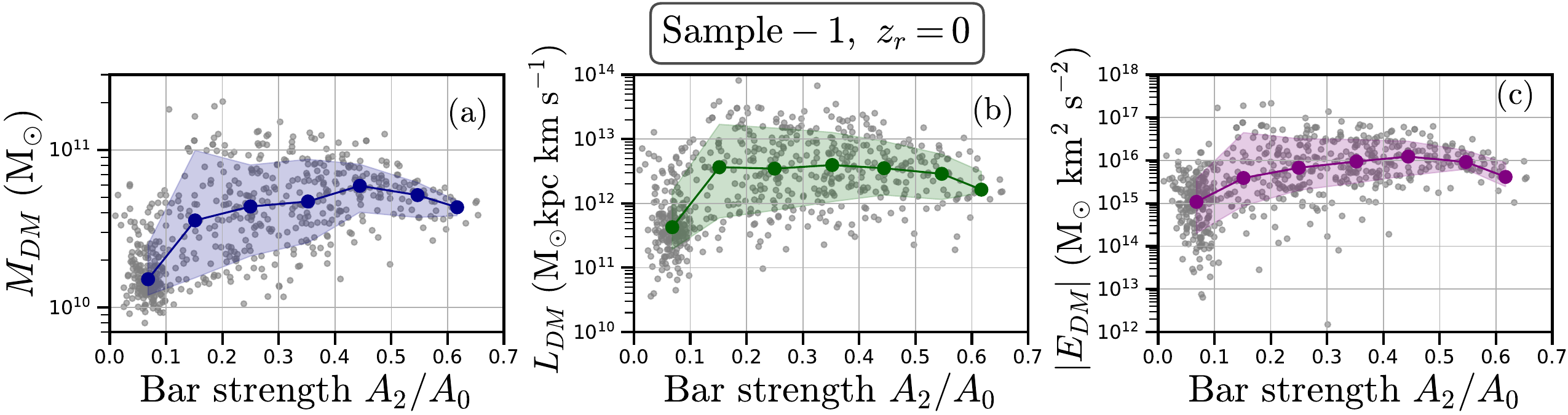} \\
\includegraphics[width=\textwidth]{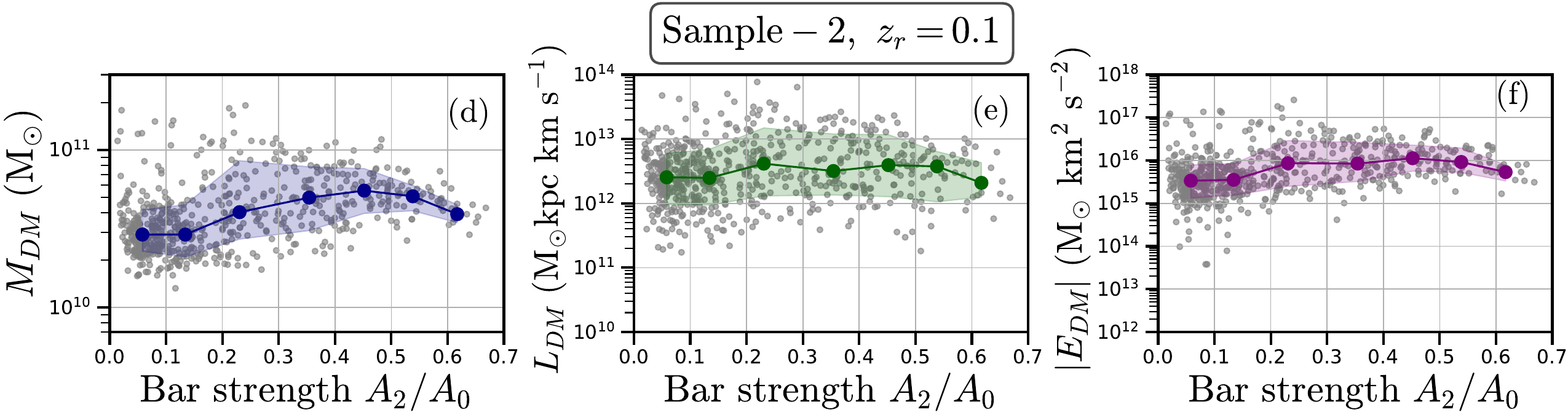}\\
\includegraphics[width=\textwidth]{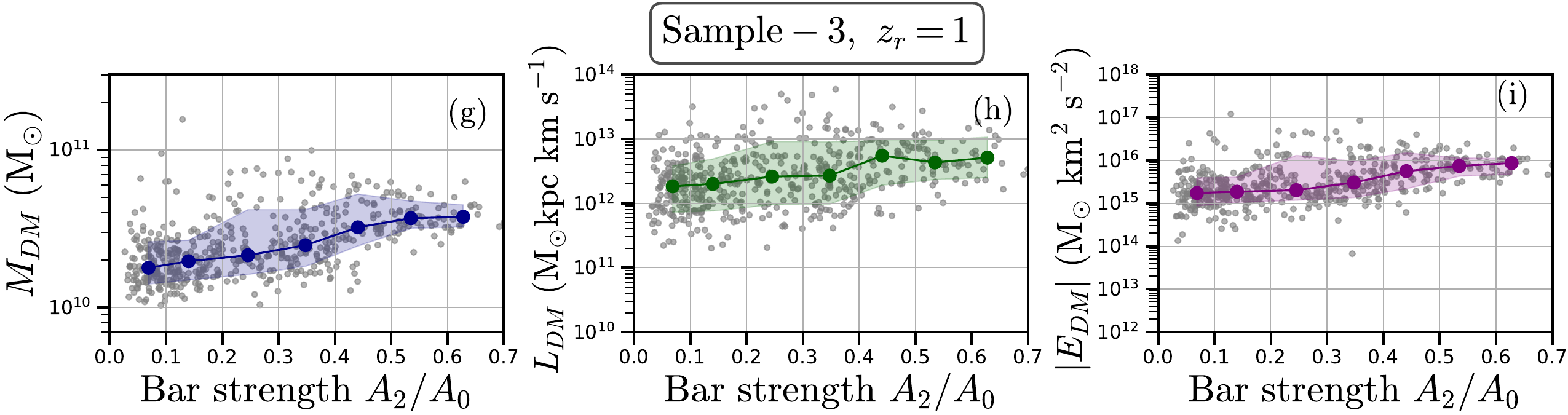}
\caption{{\bf Dark matter halo mass (left column), angular momentum (middle column) and energy (right column) of the sample galaxies at redshifts $z_{r}=$0.0 (top row), 0.1 (middle row) and 1.0 (bottom row), within spherical radius 10 kpc.}  In each panel, grey points are values of each quantity divided into seven bar strength bins (Table \ref{table:bar_strength_galaxy_no}). Blue, green and purple lines trace the median values in each bar strength bin, and the shaded regions show the 16$^{th}$ and 84$^{th}$ percentile of the sample. Each bar strength interval consists of DM halos with a wide range of mass, angular momentum and energy, however, the scatter decreases at high redshifts.  }
\label{fig:DM_halo_properties}
\end{figure*}

In Figure \ref{fig:DM_halo_properties} we present the underlying properties of the dark matter halo --- halo mass $M_{DM}$ (left column), halo angular momentum $L_{DM}$ (middle column) and the modulus of halo energy $ E_{DM}$ (right column) --- for the galaxies in the three samples at $z_r=0$ (top row), 0.1 (middle row) and 1 (bottom row), all calculated within $r_m=10$ kpc. The grey circles represent galaxies, and the median values of the above properties in each bar strength bin are in blue, green and purple circles with shaded regions indicating the 16 and 84 percentile of the distributions.

At $z_r=0.0$, a large number of low mass halos ($M_{DM}(r<10 \text{ kpc})<10^{10}$ M$_{\odot}$) is prominently seen in panel a of Figure \ref{fig:DM_halo_properties} that have low angular momentum (panel b) and low energy (panel c). This population brings down the median value of halo spin for the unbarred galaxies with $A_2/A_0<0.1$ (see Figure \ref{fig:appendix:selection_bias} in Appendix \ref{sec:appendix:selection_bias}). Some of the low-mass galaxy population is present in the high redshift Sample 2 ($z_r=0.1$, middle row) and Sample 3 ($z_r=1.0$, bottom row), which decreases the median DM mass of unbarred galaxies to lower values compared to the barred galaxies (see panels d and g). However, the DM angular momentum and DM energy distribution seem uniform over the entire range of bar strength and less affected by the low-mass galaxy population.  In conclusion, it is essential to construct galaxy samples by imposing limits on DM mass to avoid inconsistent samples and misleading correlations between halo spin and bar strength at different redshifts. 

In Section \ref{sec:halo_spin_similar_mass}, we have seen the halo spin -- bar strength anti-correlation at $z_r=0$ for galaxies with similar DM masses, where the DM mass range is fixed within a radius of $r_m=5$ and 10 kpc. This raises the question of whether the spin of the DM halo within a radius linked to the stellar disk mass is relevant for studying the connection between bars and DM halos. While this radius accommodates our galaxy samples' varying stellar masses and sizes, it does not account for similar DM masses. We also need to test if the anti-correlation between halo spin and bar strength persists when halo spin is measured at large radii, unrelated to the stellar disk, since halo spin at the virial radius should not affect/be affected by internal stellar dynamics. If a slight anti-correlation between halo spin at the virial radius and bar strength is found, it will support our conclusion from Section \ref{sec:origin_halo_spin}, that massive disks in massive DM halos, which host stronger bars, have lower halo spin compared to less massive disks in smaller halos.

We determine DM halo spin at 5 different radii. Two of these are $R_{\star, 60}$ and $R_{\star, 90}$, the radius at which the stellar mass reaches 60\% and 90\% of the total mass inside a spherical region (centered at disk center) of radius 30 kpc (see Appendix \ref{appendix:different_radii}). The other three radii are $R_{200}, 0.15 R_{200}$ and $0.05 R_{200}$, where $R_{200}$ is the virial radii of the halo (see Appendix \ref{appendix:different_radii}). 
We find a weak anti-correlation between the median halo spin and the median bar strength at low redshifts. The anti-correlation is relatively stronger at smaller radii close to the stellar disk, while at large radii, it weakens and the variation of median halo spin is constant across the entire range of bar strength. At high redshifts there is no anti-correlation and the median halo spin is nearly constant across the entire range of bar strength. For more details, see Appendix \ref{sec:appendix:all_spin}.

\subsection{Convergence}\label{sec:convergence}
We investigate the issue of numerical relaxation following \citet{Power2003}. We find that numerical convergence is satisfied for all DM halos for $r_m > 2$ kpc. We consider only those DM halos which fulfil numerical convergence. 

More than 77\% (99\%) of galaxies in Sample 1 (3) show numerical convergence for $r_m=R_{\star, 60}$ at $z_r=0.0$ ($z_r=1.0$). While we estimate the DM halo properties at $r_m=R_{\star, 60}$, our new galaxy samples consist of 460 galaxies at $z_r=0.0$, 626 galaxies at $z_r=0.1$ and 505 galaxies at $z_r=1.0$. The new galaxy sample is sufficiently large to investigate DM halo properties as a function of the bar strength of the galaxies. Among the different radii we estimate in the entire article (namely, $r_m=5 \text{ kpc, } 10\text{ kpc, } R_{\star, 60}, R_{\star, 90}, R_{200}, 0.15 R_{200}$ and $0.05 R_{200}$), only for $r_m=R_{\star, 60}$ numerical convergence is an issue for 23\% (1\%) of the galaxies in the original sample at $z_r=0$ ($z_r=1.0$).

The galaxies that do not satisfy the convergence criteria at $r_m=R_{\star, 60}$ are mostly the low-mass galaxies that are not included in the galaxy samples in Sections \ref{sec:sample1_redshift0} and \ref{sec:highz_sample23}, where we use $r_m=5$ and 10 kpc and use massive galaxies with $V_{c, DM}(r<r_{m})>140$ km s$^{-1}$.  Low-mass galaxies have small sizes. For low-mass galaxies, $R_{\star, 60}$ are smaller compared to the high-mass galaxies (see Figure \ref{appendix:fig:R30_R60_R90_rconv} in Appendix \ref{appendix:different_radii}). We exclude these low-mass galaxies while estimating halo spin for $r_m=R_{\star, 60}$ at $z_r=0.0,$ 0.1 and 1.0 in Figure \ref{appendix:median_halo_spin_bar_strength} (panels c, f and i ) in Appendix \ref{sec:appendix:all_spin}. For other values of $r_m$ (i.e., 5 kpc, 10 kpc, $R_{\star, 90}$, $R_{200}$, $0.15 R_{200}$ and $0.05 R_{200}$) the low-mass galaxies satisfy the convergence criteria. For more details on how we avoid the effects of numerical relaxation at small radii in our samples, see Appendix \ref{appendix:different_radii}.

\section{Discussion} \label{sec:discussion}
\noindent
{\it Bar strength and DM halo spin anti-correlation at $z_r=0$:} Formation of a stellar bar depends on the stellar-to-dark matter mass ratio in the disk (\citet{Bland-Hawthorn.et.al.2023}, also seen for the galaxy samples in this article; see panel e in Figure \ref{fig:spin_evolution_highz} and Appendix \ref{appendix:baryon_fraction}), the kinematic coldness of the disk, gas fraction in the disk \citep{Ostriker.and.Peebles.1973, Ansar.et.al.2023b}, strong satellite interaction events that trigger bar instability \citep{Zana.et.al.2018a, Zana.et.al.2018b, Rosas-Guevara.et.al.2022, Izquierdo-Villalba.et.al.2022, Ansar.et.al.2023b, Rosas-Guevara2024} and presence or absence of AGN activity \citep{Zhou.et.al.2020, Irodotou.et.al.2022, Kataria.M.2023}.   In this article, we show an anti-correlation between median DM halo spin and median bar strengths of the TNG50 galaxies at $z_r=0$, having similar DM mass at small radii close to the stellar disk ($r_m=5$ and 10 kpc). On tracing back the $z_r=0$ galaxies to high redshifts, we find that the anti-correlation vanishes ($z_r=1$). For an independent sample of high redshift ($z_r=1$) galaxies having similar DM mass within $r_m=10$ kpc, there is a weaker anti-correlation between the median DM halo spin and median bar strength in the range $0.2<A_2/A_0<0.7$. However, the median halo spin of galaxies with $A_2/A_0<0.1$ and $A_2/A_0>0.5$ is similar. There seems to be a more complicated relation between the halo spin and the bar strength of galaxies at high redshifts. \\

\noindent
{\it The cause of the anti-correlation between bar strength and DM halo spin:} The origin of the anti-correlation between halo spin and bar strength at low redshift results from the difference in the cosmological evolution of the DM mass and DM angular momentum of barred and unbarred galaxies. Bars from earlier in massive galaxies that form disks earlier in during evolution \citep{Khoperskov.et.al.2023}. If we consider barred and unbarred galaxies within a range of DM mass (within fixed radius $r_m$), the unbarred galaxies have lower DM mass than barred galaxies (for $z_r\geq0$; Figure \ref{fig:spin_angularmom_mass_strength_evolution_highz}). Similarly, the rise of DM angular momentum in unbarred galaxies overtakes the rise of DM angular momentum in barred galaxies at high redshifts, such that the net DM angular momentum for barred galaxies is lower than the unbarred galaxies at $z_r=0$ (see Figure \ref{fig:spin_angularmom_mass_strength_evolution_highz}). The combination of these two phenomena leads to high halo spin of the unbarred galaxies compared to the barred galaxies at $z_r=0$.  However, we do not expect the bar's influence in the galaxies' outer regions. Still, we observe a slight anti-correlation between halo spin and bar strength at the outer radii (also seen in \citep{Rosas-Guevara.et.al.2022}). This indicates that the bar may not be responsible for the anti-correlation between bar strength and halo spin in the outer regions of DM halos. \\

\noindent
{\it Does the bar play any role?} As the stellar bars grow in strength, their pattern speed decreases due to the loss of angular momentum of stars as they shift from disk to the bar \citep{Debattista.and.Sellwood.2000}. The loss of angular momentum of the bar leads to a gain in angular momentum in the outer stellar disk or a gain in angular momentum of the surrounding DM halo through resonance interactions \citep{Athanassoula.2002, Athanassoula.and.Misiriotis.2002, Petersen2016, Collier.et.al.2019a, Collier.et.al.2019b, Petersen2019, Li.et.al.2023a, Li.et.al.2023b}. Simultaneously, we notice a decrease in the stellar disk angular momentum (within $r_m<10$ kpc), even though the stellar mass within the same volume is seen to increase due to an increased number of stars. The DM gains angular momentum during the process, which results from two phenomena --- 1. an increase in DM mass within the volume ($r_m=10$ kpc), which maybe due to adiabatic contraction \citep{Blumenthal.et.al.1986, Kazantzidis.et.al.2004, Gnedin.et.al.2004} and 2. the transfer of angular momentum from the bar region to the DM halo. Here, we have not quantified the angular momentum contribution to the DM halo from the two processes as that will involve analysis of DM particle orbits having a similar frequency as the corotation resonance (CR), Inner Lindblad Resonance (ILR) and Outer Lindblad Resonance (OLR) \citep{Athanassoula.2002, Athanassoula.and.Misiriotis.2002}. The time resolution of the TNG50 data set is not high enough to conduct the analysis, which is possible for constrained galaxy simulations (for example, \citet{Collier.et.al.2019a, Collier.et.al.2019b}). Moreover, identifying the bar--halo angular momentum transfer in a cosmologically evolving galaxy is tricky, where multiple other external phenomena, (for example, gas in-fall/out-flow; satellite interactions \citet{Rodriguez-Gomez.et.al.2017}) and internal phenomena (stellar feedback and AGN activity) can as well modify the halo and disk angular momentum. Nevertheless, we find more evidence of interaction between the stellar bar and the DM halo. We observe shadow DM bars that arise in the DM density distribution once the bar grows strong and lengthens with evolution. The shadow DM bars are missing in unbarred and weakly barred galaxies (Section \ref{sec:dark_bar}). \\

\noindent
{\it Bias due to sample selection and the effect of resolution:} The choice of the sample of galaxies and the radius $r_m$ at which the halo spin is measured, both affect the median halo spins. In TNG50, low mass galaxies ($M_{\star}/M_{\odot}<10^{10}$, $M_{DM}/M_{\odot}<10^{11}$) without any bar ($A_2/A_0<0.2$) seems to have low halo spin compared to high mass galaxies (see Figure \ref{appendix:fig:median_halo_spin_strength_samehalomass3} in Appendix \ref{appendix:sec:decrease_of_halo_spin}). However, for low mass galaxies in TNG50 we also have to be cautious about the resolution at the central regions of these galaxies \citep{Ansar.et.al.2023}. In our analysis, we only consider the galaxies that do not suffer from numerical convergence inside radius $r_m$, at which we estimate halo spin (Section \ref{sec:convergence}). Another disk component that we have not considered in our sample selection is the bulge mass and bulge spin. Bar formation is delayed for galaxies with massive bulges and even more delayed for high-spinning bulges. Although, compared to the DM halo, the bulge plays a minor role in angular momentum exchange \citep{Li.et.al.2023b}. 

\section{Conclusion} \label{sec:conclusion}
We find an anti-correlation between bar strength and DM halo spin for the massive galaxies ($M_{\star}>10^{10}$ M$_{\odot}$) in the TNG50 simulations at low redshifts ($z_r\leq 0.1$). The anti-correlation weakens and is more complex at high redshifts ($z_r=1$). At low redshifts, galaxies hosting strong bars have DM halos of lower spin, lower angular momentum and lower specific angular momentum than unbarred galaxies. The anti-correlation can be explained by the differences in the evolution of the DM mass and DM angular momentum of the barred and unbarred galaxies. We observe a probable role of angular momentum exchange between the stellar bar and the DM halo that affects the spin in the central regions of DM halos. The quantification of the angular momentum transfer from the stellar disk to the DM halo through the asymmetric bar needs to be investigated with higher-resolution cosmological hydrodynamical simulations.

\begin{acknowledgments}
We thank the anonymous reviewer whose suggestions have improved this article. We thank the IllustrisTNG Collaboration for making the TNG simulation data public and to be used widely by the community. The IllustrisTNG simulations were undertaken with compute time awarded by the Gauss Centre for Supercomputing (GCS) under GCS Large-Scale Projects GCS-ILLU and GCS-DWAR on the GCS share of the supercomputer Hazel Hen at the High Performance Computing Center Stuttgart (HLRS), as well as on the machines of the Max Planck Computing and Data Facility (MPCDF) in Garching, Germany. MD acknowledges the support of the Science and Engineering Research Board (SERB) CRG grant CRG/2022/004531 for this research. 
\end{acknowledgments}

\software{astropy \citep{2013A&A...558A..33A,2018AJ....156..123A}, numpy \citep{harris2020array}, matplotlib \citep{Hunter:2007}}

\appendix

\section{ Halo spin in massive and low-mass galaxies }\label{appendix:sec:decrease_of_halo_spin}
\begin{figure*}
\centering
\includegraphics[width=0.75\textwidth]{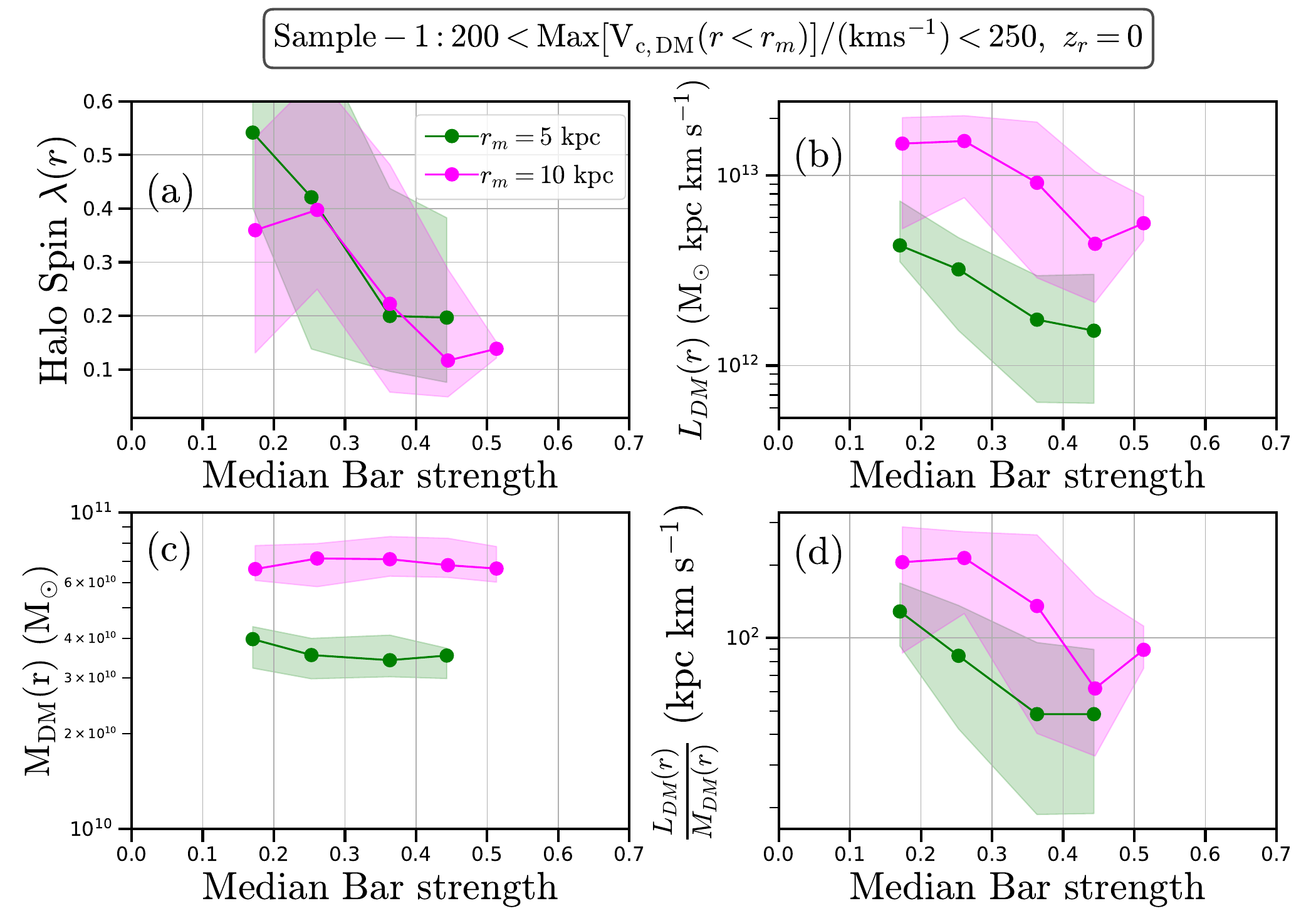}
\caption{{\bf The anti-correlation between bar strength and dark matter halo spin at $z_r=0.0$.} Same as Figure \ref{fig:median_halo_spin_strength_samehalomass}, in the circular velocity range $200< \rm Max\left[V_{c, DM}(r<r_{m})\right]/(km s^{-1})<250$. }
\label{appendix:fig:median_halo_spin_strength_samehalomass2}
\end{figure*}

We study the DM halo spin for comparable DM mass galaxies within the velocity range of $200< \rm Max\left[V_{c, DM}(r<r_{m})\right]/(km s^{-1})<250$ for $r_{m}=5$ and 10 kpc. We find 0, 6, 21, 23, 33, 8 and 0 number of galaxies for $r_{m}=10$ kpc for the seven bar strength bins in Table \ref{table:bar_strength_galaxy_no} and similarly for $r_{m}=5$ kpc we have 0, 10, 19, 19, 17, 3 and 0 galaxies for each of the seven bar strength bins. We avoid low statistics by considering the bar strength intervals with $\geq 6$ galaxies. In Figure \ref{appendix:fig:median_halo_spin_strength_samehalomass2}, we present the halo spin (panel a), halo angular momentum (panel b), halo mass (panel c) and halo specific angular momentum (panel d) for the above sample of barred and unbarred galaxies.

\begin{figure*}
\centering
\includegraphics[width=0.75\textwidth]{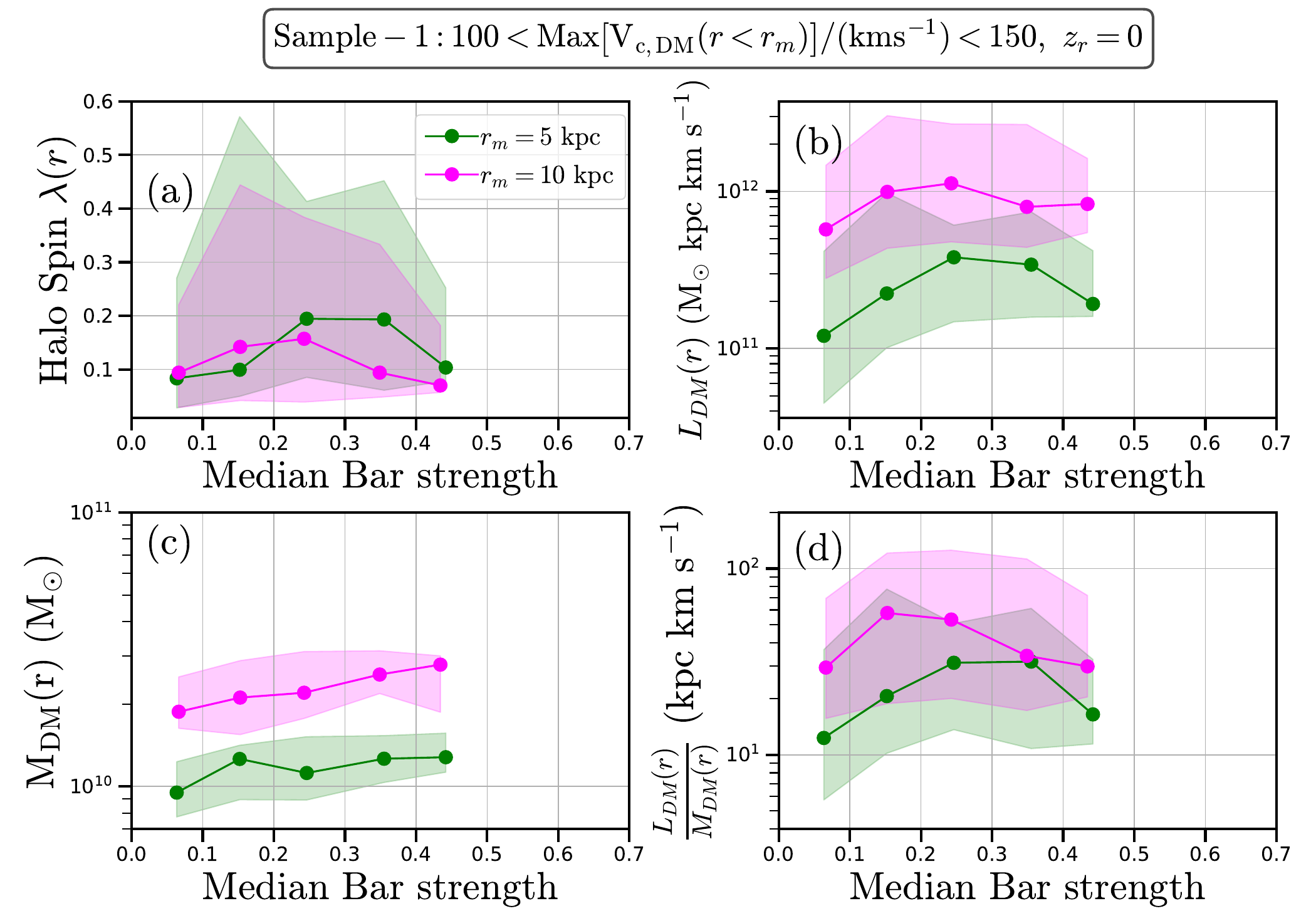}
\caption{{\bf Similar to Figure \ref{fig:median_halo_spin_strength_samehalomass}}, in the circular velocity range $100< \rm Max\left[V_{c, DM}(r<r_{m})\right]/(km s^{-1})<150$. }
\label{appendix:fig:median_halo_spin_strength_samehalomass3}
\end{figure*}

We also check the velocity range $100< \rm Max\left[V_{c, DM}(r<r_{m})\right]/(km s^{-1})<150$ with the low mass galaxies. However, we do not have uniform mass distribution for each of the bar strength bins for the low mass systems. We find 72, 23, 38, 28, 8, 3 and 3
galaxies at $r_{m}=5$ kpc and 80, 27, 36, 22, 7, 2 and 2
 galaxies for $r_{m}=10$ kpc. Again, we only consider the bar strength intervals with $\geq 6$ galaxies. In Figure \ref{appendix:fig:median_halo_spin_strength_samehalomass3} we present the halo spin of the lowest mass galaxies for different bar strength intervals.

\section{Galaxies with unequal DM mass causing selection bias}\label{sec:appendix:selection_bias}

\begin{figure*}
\centering
\includegraphics[width=\textwidth]{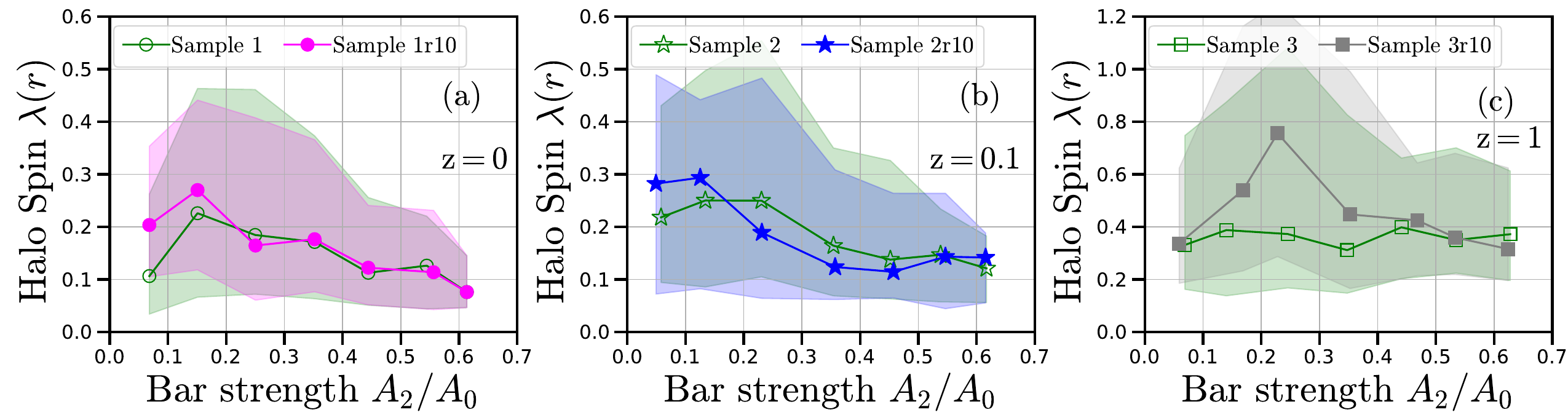}
\caption{{\bf Incorrect sample selection can lead to unclear correlations in DM halo spin.} The Figure shows the comparison of the median halo spin (within $r_m=10$ kpc) for different galaxy samples at three redshifts $z_r=0$ (panel a), 0.1 (panel b) and 1 (panel c).  The Samples 1r10 (solid circles), 2r10 (solid stars) and 3r10 (solid squares) consist of galaxies with similar DM mass following the criterion: $140< \rm Max\left[V_{c, DM}(r<r_{m})\right]/(km s^{-1})<200$ and is a subset of the larger Samples 1, 2 and 3 (hollow circles, stars and squares in green) from Table \ref{table:bar_strength_galaxy_no} that do not have any constrain on DM mass, as shown in panels a, d and g in Figure \ref{fig:DM_halo_properties}. }
\label{fig:appendix:selection_bias}
\end{figure*}

Here, we investigate the difference between the median halo spin calculated from two kinds of galaxy samples. The first sample consists of galaxies with similar DM mass within $r_m=10$ kpc, for example, the Samples 1r10 (Section \ref{sec:halo_spin_similar_mass}), 2r10 and 3r10 (Section \ref{sec:highz_sample23}). The second sample consists of galaxies without any constraints on DM mass, for example, the entire Sample 1, 2 and 3 (Section \ref{sec:barred_unbarred_sample}, Table \ref{table:bar_strength_galaxy_no}). We determine the halo spin of galaxies in all samples within $r_m=10$ kpc. Figure \ref{fig:appendix:selection_bias} shows the median halo spins for the six galaxy samples mentioned above at three redshifts $z_r=0$ (panel a), 0.1 (panel b) and 1 (panel c). The median halo spin for the unbarred galaxies in Sample 1r10 is higher than in Sample 1, as low-mass unbarred galaxies with $A_2/A_0<0.2$ in Sample 1 decrease the median halo spin. Deviations between the median halo spin for Sample 2r10 and Sample 2 (panel b), and Sample 3r10 and Sample 3 (panel c) are also due to the distribution of DM masses. Constructing galaxy samples of similar DM halo mass is important to study correlations between halo spin and bar strengths.

\section{Estimation of $R_{\star, 60}$, $R_{\star, 90}$ and $R_{200}$} \label{appendix:different_radii}
The disk interacts with the DM halo through the stellar bar by transferring angular momentum from the disk to the inner DM halo. We aim to estimate the DM halo properties at scale lengths closely associated with the stellar disk. 
We estimate the radius of the 30\%, 60\% and 90\% stellar mass within a spherical region of radius 30 kpc from the disk center, $ R_{\star,30}$, $ R_{\star, 60}$ and $ R_{\star, 90}$.  

\begin{figure*}
\centering
\includegraphics[width=0.4\textwidth]{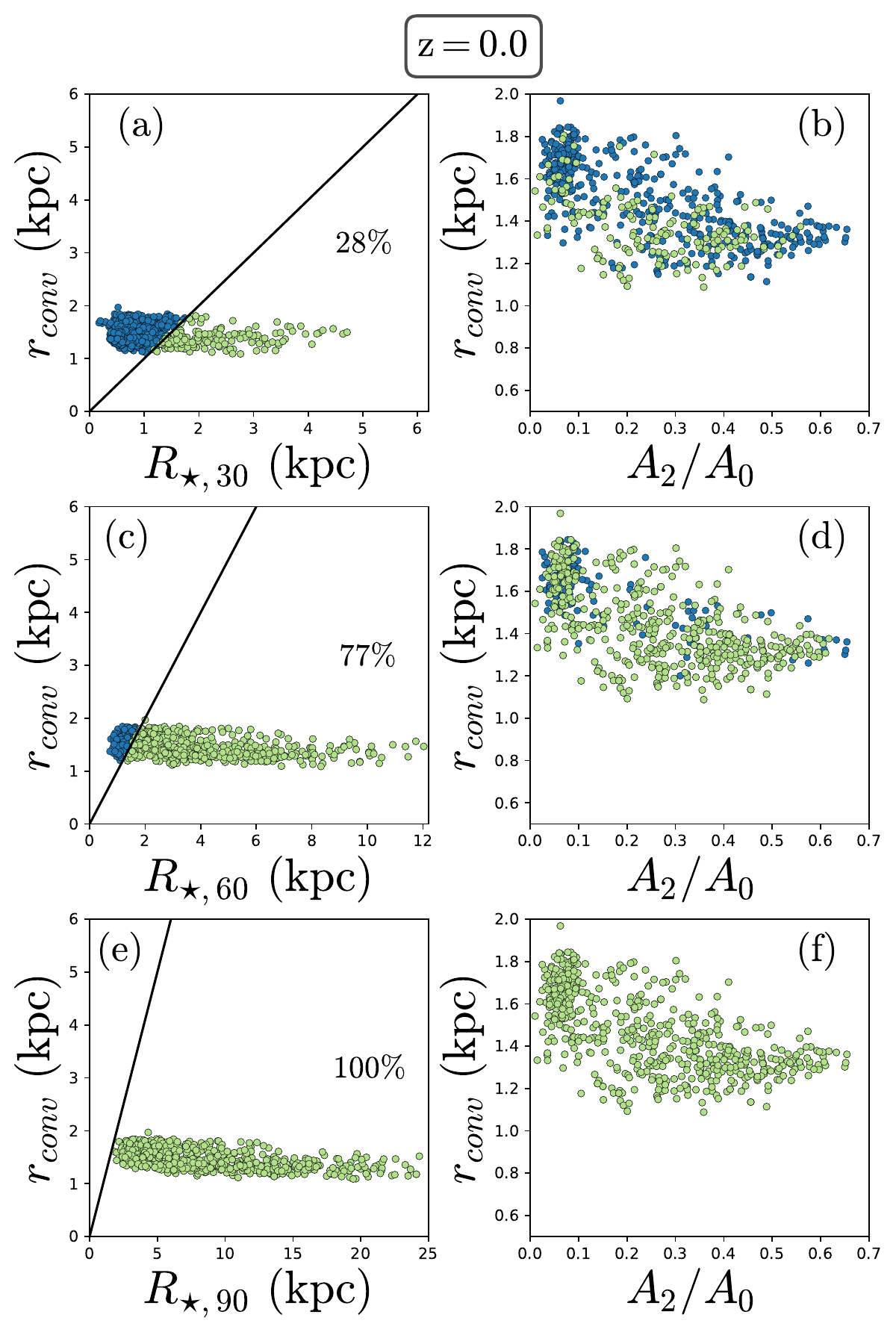} 
\includegraphics[width=0.4\textwidth]{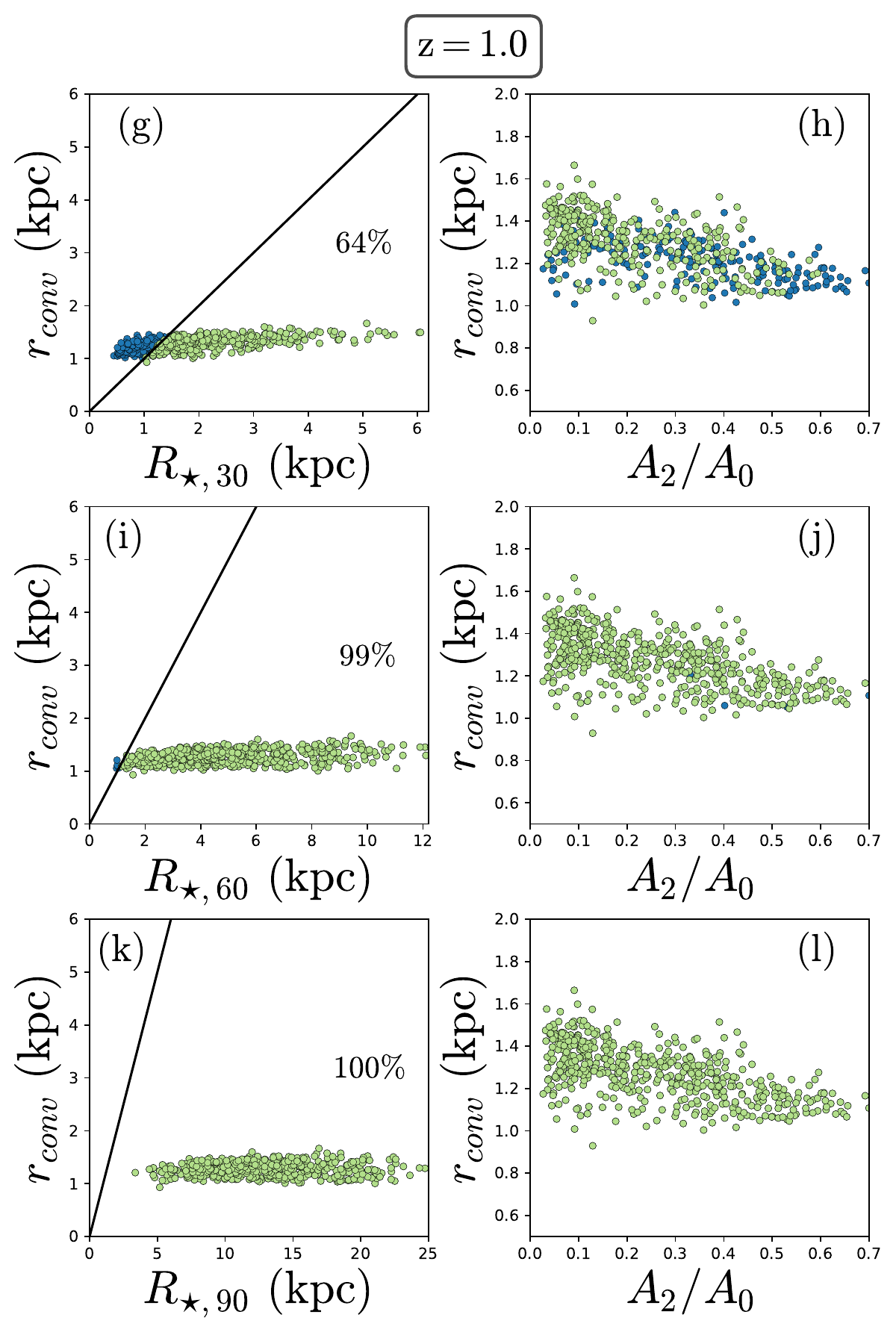}
\caption{{\bf Comparison of $R_{\star,60}$ and $R_{\star,90}$ with the radius of convergence $r_{conv}$ at redshift $z_r=$ 0.0 (first and second column) and 1.0 (third and fourth column).} First and third column shows $R_{\star,\rm X}$ ($\rm X=60$ and 90) versus $r_{conv}$ (see text in Appendix \ref{appendix:different_radii}), and second and fourth column shows $r_{conv}$ versus bar strength $A_2/A_0$ for the same galaxies. The green circles represent galaxies with $R_{\star,\rm X}>r_{conv}$  and the blue circles for galaxies with $R_{\star,\rm X}<r_{conv}$. The percentage of green circles is shown in the different panels. As we move from $R_{\star,60}$ to $R_{\star,90}$ the fraction of galaxies satisfying the numerical convergence ($R_{\star,\rm X}>r_{conv}$) increases to 100\% for $R_{\star, 90}$ at all redshifts.}
\label{appendix:fig:R30_R60_R90_rconv}
\end{figure*}

Most of the disks in our sample are well within the sphere of 30 kpc radius. Hence it is reasonable to estimate the radius containing 30\%, 60\% and 90\% of the stellar mass within a sphere of radius 30 kpc. Some of the disks in our sample have small values of $ R_{\star,30}$ and $ R_{\star,60}$, and estimation the DM halo properties at small radii can suffer from issues of numerical relaxation. Following the method presented in \citet{Power2003}, we estimate the radius that bounds the volume inside which the two-body relaxation time scale $t_{relax}$ is smaller than the age of the Universe $t_{age}$. The radius at which $t_{relax}=t_{age}$, can be adopted as the radius of convergence ($r_{conv}$) of the inner region of the DM halo. $r_{conv}$ depends on the number of DM particles ($N(<r)$), the mean density ($\bar{\rho}(r)$) and the critical density of the Universe at redshift $z_r$ ($\rho_{crit}(z_r)=3 H^2(z_r)/8 \pi G$), and is a solution to the following equation:
\begin{equation}
    \frac{t_{relax}}{t_{age}}= \frac{\sqrt{200}}{8} \frac{N(r)}{\ln N(r)} \left(\frac{\bar{\rho}(r)}{\rho_{crit}(z_r)}\right)^{-1/2}=1  \text{     .}
\end{equation}
We solve the above equation for the radius of convergence $r_{conv}$ in each of the galaxy DM in our samples and compare $r_{conv}$ with $ R_{\star,30}$, $ R_{\star,60}$ and $ R_{\star,90}$. In Figure \ref{appendix:fig:R30_R60_R90_rconv}, where we show their comparison at two redshifts $z_r=0.0$ (first and second column for Sample-1) and 1.0 (third and fourth column for Sample-3). The first and third column show the comparison of $R_{\star,\rm X}$ ($\rm X =30, 60$ and 90) and  $r_{conv}$ and the black line marks the boundary of $R_{\star,\rm X}=r_{conv}$. The second and fourth columns show how $r_{conv}$ varies with the bar strength of each galaxy from our samples. From visual inspection, the galaxies in green and blue seem to be uniformly distributed throughout the bar strength range. In Figure \ref{appendix:fig:R30_R60_R90_rconv} in all panels, each circle represent a galaxy from our Samples 1 and 3. 

The convergence criteria is fulfilled by the galaxies shown in green for which $R_{\star,\rm X}> r_{conv}$ (percentage of galaxies shown inside panels), and not by the galaxies shown in blue where $R_{\star,\rm X}< r_{conv}$. At $R_{\star,30}$, less number of galaxies follow the above criteria compared to $R_{\star,90}$. To ensure numerical convergence for the majority of the sample, we choose galaxies with $R_{\star,\rm X}> r_{conv}$ and estimate the halo properties at $R_{\star, 60}$ and $R_{\star, 90}$. This ensures that we have a sufficient number of galaxies in our samples to estimate halo properties and also avoid issues related to the numerical relaxation of DM particles.

Till now we have focused on length scales associated with the stellar disk that can be useful to measure DM halo spin. However, we also want to study the halo spin at different halo radii, for example, the virial radius $R_{200}$. $R_{200}$ is defined as the radius inside which the average density is 200 times the critical density of the Universe:
\begin{equation}
    M_{200}=\frac{4}{3}\pi R^3_{200} \times (200 \rho_{crit}(z_{r}) )
\end{equation}
where, $M_{200}$ is the virial mass and $\rho_{crit}(z_{r})= 3 H^2(z_{r})/8\pi G$ is the critical density at redshift $z_{r}$. We estimate the average density at different radii and a match with $200\times \rho_{crit}(z_{r})$ gives $R_{200}$. To study the effect of the disk we measure the halo spin at $R_{200}$ away from the influence of the stellar disk, and at radii closer to the disk - $0.15 R_{200}$ and $0.05R_{200}$.

\section{Median halo spin for galaxies after relaxing the fixed DM mass criterion}\label{sec:appendix:all_spin}

We calculate the halo spin at multiple radii of the DM halo, namely $r_m=5,$ and 10 kpc in Section \ref{sec:halo_spin_similar_mass} and \ref{sec:highz_sample23}. We also determine the halo spin at $R_{\star, 60}$ and $R_{\star, 90}$, the radius at which the stellar mass reaches 60\% and 90\% of the total mass inside a spherical region (centered at disk center) of radius 30 kpc. To estimate $R_{\star, X}$, we choose a maximum radius of 30 kpc as all of the galaxies in our sample are well within the 30 kpc radius. In addition, we estimate the halo spin at the virial radius $ R_{200}$, far from the central disk and close to the disk at $0.15 R_{200}$ and $0.05 R_{200}$, where we expect a greater impact of the disk on the surrounding DM particles. We present the detailed method of estimation of each radius in Appendix \ref{appendix:different_radii}.

\begin{figure*}
\centering
\includegraphics[width=\textwidth]{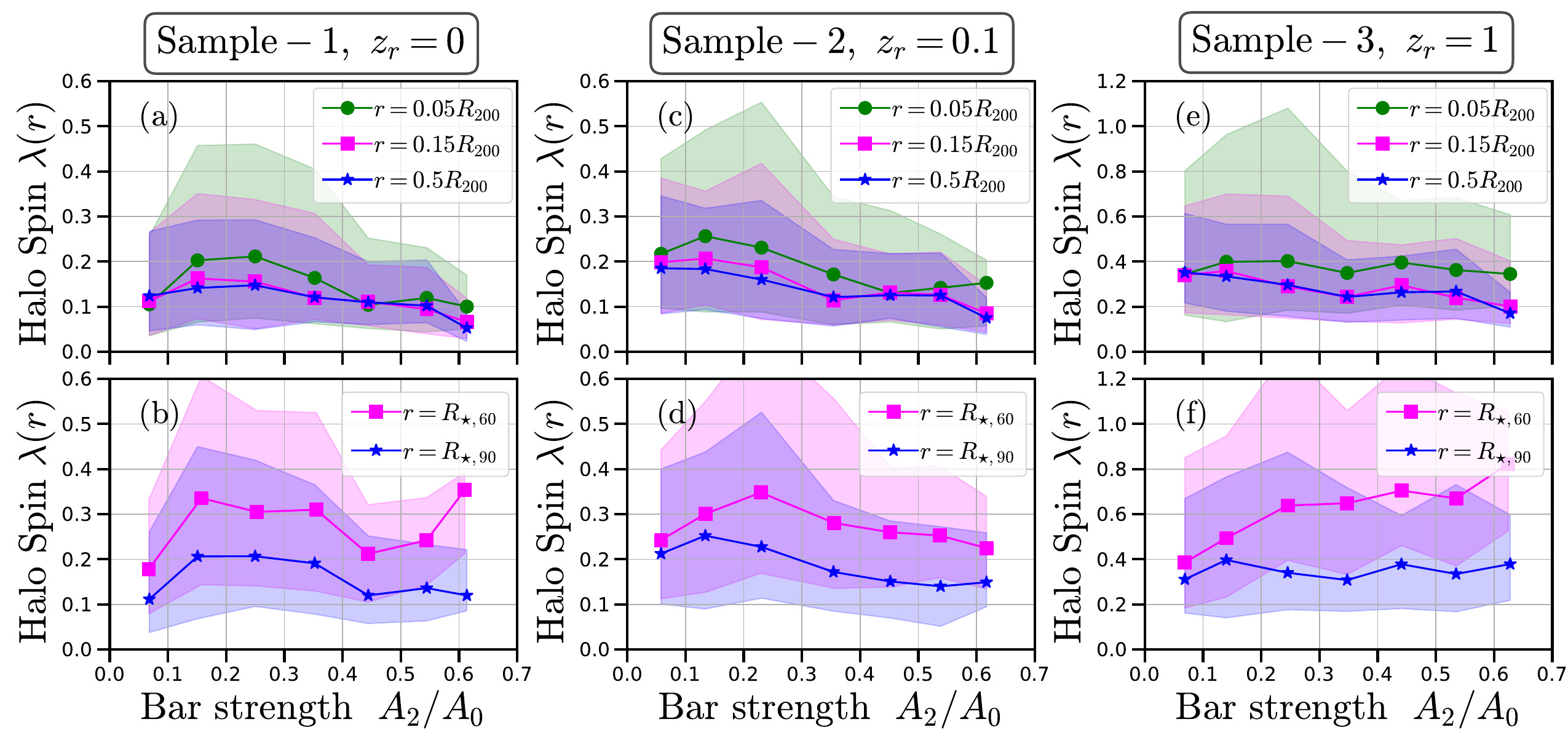}
\caption{{\bf The connection between bar strength and dark matter halo spin at redshift $z_{r}=$0.0 (top row), 0.1 (middle row) and 1.0 (bottom row).} Each panel shows the median halo spin and median bar strength (circles, squares and ``$\star$") for each of the bar strength bins from Table \ref{table:bar_strength_galaxy_no}, for different radii. In all panels, green circles are for the smallest radii, magenta squares are for the next radii and blue stars are for the largest radii, with the shaded regions of similar colors bounded by the 16$^{th}$ and 84$^{th}$ percentile curves of halo spin distribution. Left column shows halo spin at $r=5$, 10 and 100 kpc; middle column for $r=0.05 R_{200}, 0.15 R_{200}$ and $0.5 R_{200}$; the right column for $r=R_{\star, 60}$ and $R_{\star, 90}$ (see Section \ref{sec:halo_spin_all_mass}). At low redshift and small radii, the median halo spin decreases with increasing bar strength. The shape of the shaded regions highlights the skewed nature of the spin distribution in each bar strength bin. }
\label{appendix:median_halo_spin_bar_strength}
\end{figure*}

In Figure \ref{appendix:median_halo_spin_bar_strength} we present the median halo spins of the galaxies in different bar strength bins for the three samples from Table \ref{table:bar_strength_galaxy_no} at redshift $z_{r}=0.0$ (left column), 0.1 (middle column) and 1.0 (right column), and at different radii $r_m=0.05 R_{200}, 0.15 R_{200} $ and $0.5R_{200}$ (top row) and $r_m=R_{\star, 60}$ and $R_{\star, 90}$ (bottom row). The median halo spin for different bar strength bins are in solid circles, squares and stars, and the shaded regions show the $16^{th}$ and $84^{th}$ percentile of the population. The different colors in each panel represent the different radii at which we estimate halo spin, with blue for the largest radii, magenta for the intermediate radii and green for the smallest radii.

At $z_r=0.0$ (Figure \ref{appendix:median_halo_spin_bar_strength}, top row), except at the very low bar strength interval ($0<A_2/A_0<0.1$), we observe an overall similar trend (as in Figure \ref{fig:median_halo_spin_strength_samehalomass}) of decrease in halo spin with higher bar strength for all the radii in different panels. In general, for larger radii, the median halo spin decreases for nearly all the bar strength intervals. The spread in the halo spin distribution (shaded regions) is broader in the low bar strength intervals compared to the high bar strength intervals (similar to Figure \ref{appendix:median_halo_spin_bar_strength}). There is a decrease in the median halo spin in the lowest bar strength interval ($0<A_2/A_0<0.1$) which is dominated by the low mass, low spin galaxies mentioned in Section \ref{sec:halo_spin_similar_mass} and \ref{sec:halo_spin_all_mass}. 

The halo spin at higher redshifts (Figure \ref{appendix:median_halo_spin_bar_strength}, middle row and bottom row), for example at $z_r=1.0$ is very different from halo spin at $z_r=0.0$, although the spin distribution at $z_r=0.1$ has more similarity with spin at $z_r=0.0$. There is significantly less number of low-mass galaxies that settle into a disk in our high redshift samples and unlike $z_r=0.0$, we observe a higher median halo spin of the unbarred galaxies at $z_r>0.0$. Overall, the halo spins for unbarred galaxies ($0<A_2/A_0<0.2$) are higher compared to the barred galaxies ($0.2<A_2/A_0<0.7$) even at $z_r=0.1$. However, the low redshift trend does not hold at high redshift. At $z_r=1.0$, the median halo spin close to the disk (green lines in panel e, and magenta and blue lines in panel f) is either similar for barred and unbarred galaxies or the trend reverses at very low radii (magenta line in panel f). At large radii, away from the disk (blue and magenta lines in panel e), the halo spin is higher for unbarred galaxies compared to barred galaxies.

The DM halo spin of the three galaxy samples at different redshifts indicates an evolutionary nature of halo spin that differs for the barred and unbarred galaxies. 

\section{Mass fraction in the disk within $R_{\star, 90}$} \label{appendix:baryon_fraction}

\begin{figure*}
\centering
\includegraphics[width=\textwidth]{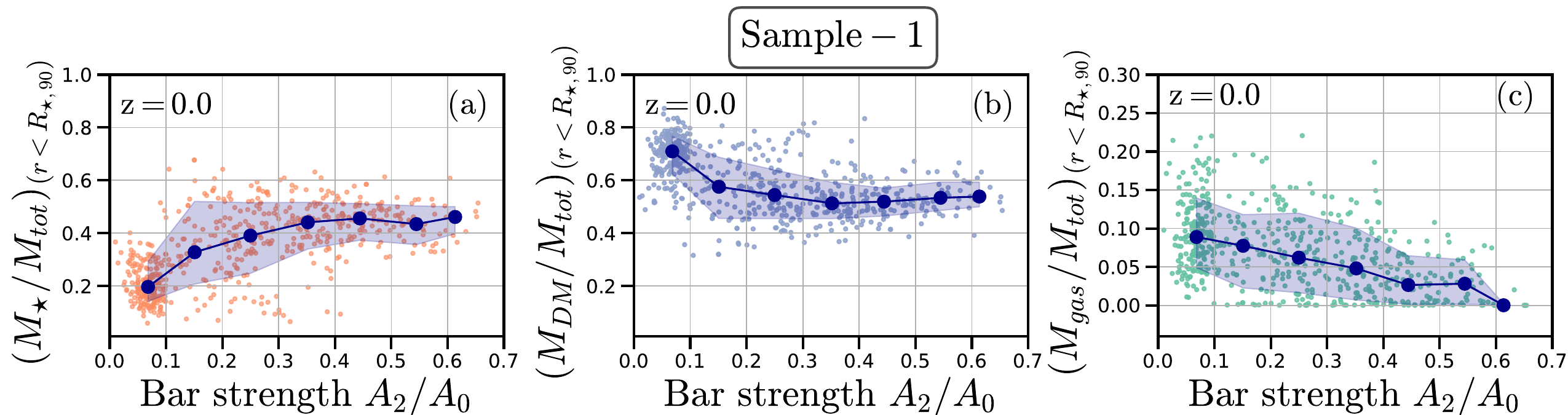} \\
\includegraphics[width=\textwidth]{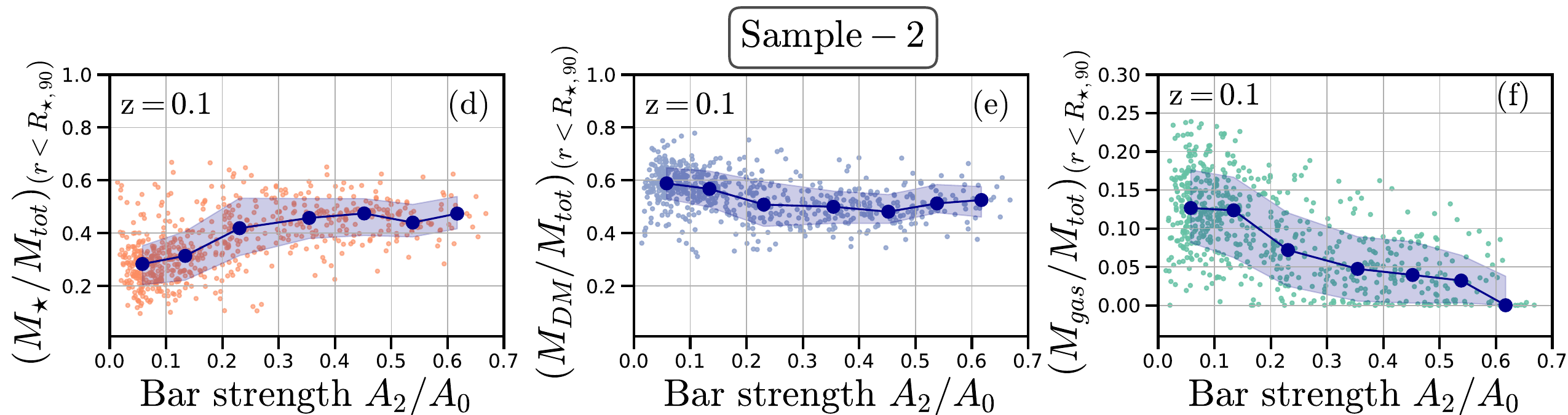}\\
\includegraphics[width=\textwidth]{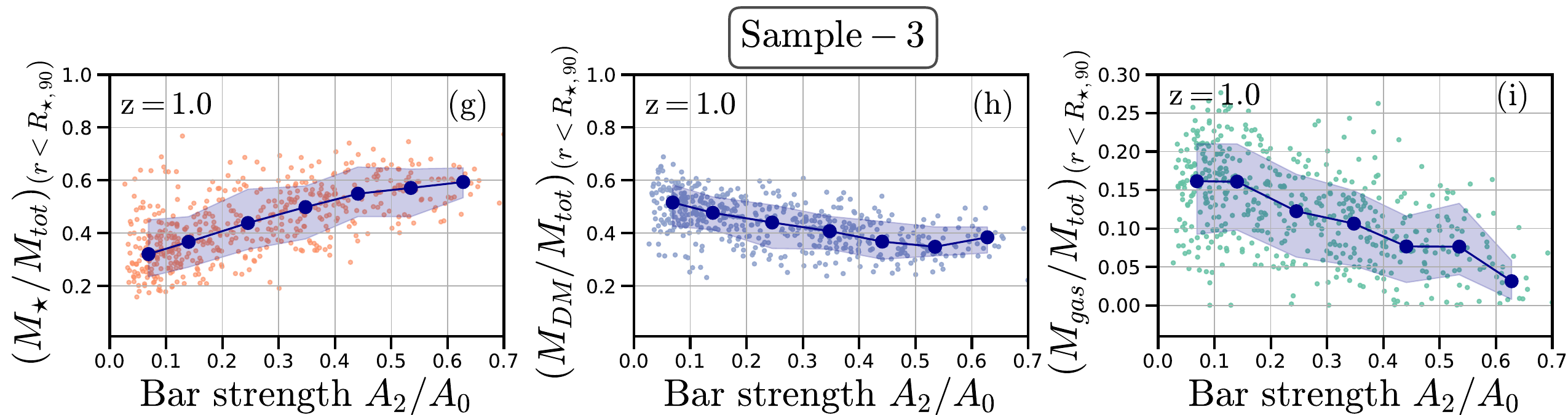}
\caption{{\bf Stellar mass fraction ($M_{\star}/M_{tot}$; left column), dark matter mass fraction ($M_{DM}/M_{tot}$; middle column) and gas mass fraction ($M_{gas}/M_{tot}$; right column) at redshift $z_{r}=$0.0 (top row), 0.1 (middle row) and 1.0 (bottom row), within spherical radius $R_{\star, 90}$ and $|z|<2$ kpc.}  In each panel, reddish/blue/green points are values of each quantity in different bar strength bins (Table \ref{table:bar_strength_galaxy_no}). Blue lines connect the median values (dark blue circles) in each bar strength bin and the shaded regions are bounded by the 16$^{th}$ and 84$^{th}$ percentile of each of the above quantities in each bar strength bin. }
\label{appendix:mass_fraction}
\end{figure*}

Figure \ref{appendix:mass_fraction} shows the stellar mass fraction ($M_{\star}/M_{tot}$; orange circles in left column), the DM mass fraction ($M_{DM}/M_{tot}$; blue circles in middle column) and the gas mass fraction ($M_{gas}/M_{tot}$; green circles in right column) within radius $r<R_{\star, 90}$ and $|z|<2$ kpc from the disk mid-plane, of all our sample galaxies at redshifts $z_r=0.0$ (top row), 0.1 (middle row) and 1.0 (bottom row). Here $M_{tot}$ is the combined mass of all components. The dark-blue big solid circles show the median mass fraction in each panel, and the shaded region is bounded by the 16$^{th}$ percentile and the 84$^{th}$ percentile curves.

As we move towards galaxies with higher bar strengths, the median value of the stellar mass fraction in the disk plane increases, the median of the DM mass fraction slightly decreases and the median gas mass fraction decreases to low values close to less than 5\%. At high redshifts, the stellar mass fraction and gas fraction are higher than the low redshift values. At low redshifts, the DM mass fraction increases making the stellar and gas mass fractions lower. All the panels in Figure \ref{appendix:mass_fraction} reflect the well-established idea that stronger bars tend to last in galaxies with higher stellar mass fractions and lower gas mass fractions \citep{Bland-Hawthorn.et.al.2023}. 

\bibliography{sample631}{}
\bibliographystyle{aasjournal}

\end{document}